\documentclass[12pt,preprint]{aastex}   

\def\feka{Fe K$\alpha$}   
\def\fekb{Fe K$\beta$}   
\def\chandra{{\it Chandra}}   
\def\xmm{{\it XMM-Newton}}   
\def\suzaku{{\it Suzaku}}   
\def\swift{{\it Swift}}   
\def\asca{{\it ASCA}}   
   
\def\rxte{{\it RXTE}}   
\def\sax{{\it BeppoSAX}}   
   
\def\rosat{{\it ROSAT}}

\def\fermi{{\it Fermi}}   
   
\def\lum{erg s$^{-1}$}   
\def\flux{erg cm$^{-2}$ s$^{-1}$}   
\def\nh{cm$^{-2}$}   
\def\arcsec{$^{\prime\prime}$}   
\def\deg{$^{\circ}$}   
   
\def\ltsima{$\; \buildrel < \over \sim \;$}   
\def\simlt{\lower.5ex\hbox{\ltsima}} 
\def\gtsima{$\; \buildrel > \over \sim \;$}   
\def\simgt{\lower.5ex\hbox{\gtsima}} 
   
\def\3c{3C~382}

\begin{document}   
   
\title{The Suzaku view of 3C~382}   
   
\author{R. M. Sambruna~$^{1}$, F. Tombesi~$^{2,3}$, J.~N. Reeves~$^{4}$, V. Braito~$^{5}$, L. Ballo~$^{6}$, M. Gliozzi~$^{1}$ and C.~S. Reynolds~$^{3}$} 
\affil{$^1$ Department of Physics and Astronomy, MS 3F3, 4400 University Drive, George Mason University, Fairfax, VA 22030}
\affil{$^2$ X-ray Astrophysics Laboratory and CRESST, NASA/Goddard Space Flight Center, Greenbelt, MD 20771, USA}
\affil{$^3$ Department of Astronomy, University of Maryland, College Park, MD 20742, USA}
\affil{$^4$ Astrophysics Group, School of Physical and Geographical Sciences, Keele University, Keele, Staffordshire ST5 5BG, UK}
\affil{$^5$ Department of Physics and Astronomy, University of Leicester, University Road, Leicester LE1 7RH, UK}
\affil{$^6$ Istituto de Fisica de Cantabria (CSIC-UC), 39005 Santander, Spain}

%
%
%

\begin{abstract}

We present a long (116~ks) \suzaku\ observation of the Broad-Line
Radio Galaxy (BLRG) \3c\ acquired in April 2007. A \swift\ BAT
spectrum in 15--200~keV from the 58-month survey is also analyzed,
together with an archival \xmm\ EPIC exposure of 20~ks obtained one
year after \suzaku. Our main result is the finding with
\suzaku\ of a broad Fe~K line with a relativistic profile consistent
with emission from an accretion disk at tens of gravitational radii
from the central black hole. The XIS data indicate emission from
highly ionized iron and allow us to set tight, albeit model-dependent,
constraints on the inner and outer radii of the disk reflecting
region, $r_{in} \simeq 10 r_g$ and $r_{out} \simeq 20 r_g$,
respectively, and on the disk inclination, $i\simeq 30$\deg. Two ionized 
reflection components are possibly observed, with similar contributions of 
$\sim$10\% to the total continuum. A highly ionized one, 
with log$\xi$$\simeq$3~erg~s$^{-1}$~cm, 
which successfully models the relativistic line and a mildly ionized one, 
with log$\xi$$\simeq$1.5~erg~s$^{-1}$~cm, which models the narrow Fe K$\alpha$ 
and high energy 
hump. When both these components are included, there is no further requirement for 
an additional black body soft excess below 2~keV.
The \emph{Suzaku} data confirm the presence of a warm
absorber previously known from grating studies. After accounting for
all the spectral features, the intrinsic photon index of the X-ray
continuum is $\Gamma_x\simeq1.8$ with a cutoff energy at $\sim
200$~keV, consistent with Comptonization models and
excluding jet-related emission up to these energies. Comparison of the
X-ray properties of \3c\ and other BLRGs to Seyferts recently observed
with \suzaku\ and BAT confirms the idea that the distinction between
radio-loud and radio-quiet AGN at X-rays is blurred. The two classes
form a continuum distribution in terms of X-ray photon index, reflection 
strength, and Fe~K line width (related to the disk emission
radius), with BLRGs clustered at one end of the distribution. This
points to a common structure of the central engine, with only a few
fundamental parameter(s) responsible for the radio-loud/radio-quiet
division. The black hole spin, and in particular its rotation compared
to the disk's, may be a key one.

\end{abstract}   
   
{\sl Subject Headings:} {Galaxies: active --- galaxies: radio --   
galaxies: individual --- X-rays: galaxies}

\section{Introduction}   
   
It is by now widely accepted that Active Galactic Nuclei (AGN),   
including radio-loud sources, are powered by accretion of the host   
galaxy gas onto a central super-massive black hole (e.g., Blandford   
1985). The rich variety of AGN phenomenology has been explained in   
terms of orientation with respect to the axis of an obscuring   
equatorial thick matter surrounding the nucleus (Antonucci 1993;  
Urry \& Padovani 1995),   
yielding type-1 sources for face-on views and type-2 objects for   
edge-on views. In addition, in radio-loud sources a relativistic jet   
is present, connecting the innermost regions near the black hole to   
the radio galaxy perifery.  The jet angle -- defined as the angle   
between the jet axis and the line of sight -- increases from blazars   
to Broad- and Narrow-Line Radio Galaxies, roughly corresponding to an   
increase of obscuration degree. As the jet angle increases, the   
importance of its emission decreases, due to beaming effects.    
   
X-ray spectroscopy is an effective tool to investigate the central
engines of AGN. Indeed, previous X-ray observations of BLRGs in the
1990s with \asca, \rxte, and \sax\ established that these sources
exhibit Seyfert-like spectra with subtle but significant differences,
i.e., flatter X-ray continua and weaker reflection features than in
radio-quiet (see Sambruna, Eracleous, \& Mushotzky 2002; Ballantyne
2007, and references therein, for a review). In particular, the
\feka\ emission line around $\sim$6.4 keV was observed to be narrower
and of lower EW than in Seyferts, and generally unresolved at the
limited resolution of the pre-\suzaku\ detectors. No clear evidence
for a relativistic accretion disk profile was found for the
\feka\ line in any BLRGs, or other RL AGN.  These results raised the
currently held scenario that some intrinsic, fundamental difference
must exist in the structure of the accretion flow between the two
classes of AGN, with RQ being dominated by standard optically thick,
geometrically thin disks and RL having radiatively inefficient,
ADAF-like, inner disks (see discussion in Eracleous et al. 2000;
Ballantyne 2007). Alternatively or in addition, a non-thermal jet
contribution was also possible.
    
The advent of \suzaku\ with its broad-band coverage in 0.3--100~keV
and improved sensitivity especially in the critical Fe~K region,
6--7~keV, has provided a golden opportunity to deepen our
understanding of the central engines of RL AGN. Our group has secured
GO observations of all the bright, nearby ($z<0.1$) BLRGs with
relatively deep, 100~ks, \suzaku\ exposures. The \suzaku\ observations
of our program targets are described in separate papers that
concentrate on one source at the time, as each classical BLRG presents
somewhat unique X-ray and multi-wavelength properties (3c~390.3,
Sambruna et al. 2009, S09 hereafter; 3C~445, Braito et al.~2011;
Reeves et al. 2010; 3C~111, Ballo et al. in prep.). In this paper, we
focus on \3c, a giant lobe-dominated radio galaxy well-known for its
soft excess and rapid X-ray variability from previous observations
(see \S~2). Besides its \suzaku\ data, we also discuss its \xmm\ EPIC
data and a \swift\ BAT spectrum (see below), both unpublished.
   
The \suzaku\ observation of \3c\ confirms the complexity of its   
spectrum and its similarity to Seyferts. Remarkably, we find   
compelling evidence in the XIS data for a {\it relativistic   
  \feka\ line profile}, with excellent constraints on the disk   
inclination and inner/outer radii (\S~5.3). This is the first time that a   
such a detailed profile is detected in a BLRG, and indeed in a radio-loud   
AGN. The implications for the structure of the central engine and   
models of jet formation are far reaching (\S~8).   
   
The paper is organized as follows. After describing the source   
properties and previous observations in \S~2, in \S~3 the data   
reduction of the new observations is presented. The \suzaku\ timing
analysis is reported in \S~4. We describe the spectral fits to the various 
datasets in \S~5 for the combined \suzaku\ and \swift\ BAT and in \S~6 for 
{\it XMM-Newton}. In \S~7 we report the summary of the  
results and the discussion following in \S~8. Finally, in Appendix A we 
report the background and calibration tests for the XIS cameras. Throughout 
this paper, a   
concordance cosmology with H$_0=71$ km s$^{-1}$ Mpc$^{-1}$,   
$\Omega_{\Lambda}$=0.73, and $\Omega_m$=0.27 (Spergel et al. 2003) is   
adopted. The energy spectral index, $\alpha$, is defined such that   
$F_{\nu} \propto \nu^{-\alpha}$. The photon index is $\Gamma=\alpha+1$.   
   
\section{\3c\ and Previous X-ray Observations}   
   
The lobe-dominated, Fanaroff-Riley II radio galaxy \3c\ ($z=0.0579$)   
has a double-lobe structure, with a clear jet in the northern lobe   
ending in a hotspot. While a counter-jet is not clear, a hotspot is   
also detected in the southern lobe, with a total size between hotspots   
at 3.89~GHz of 179\arcsec\ (Hardcastle et al. 1998). From this size, a   
lower limit to the jet inclination of $\theta=15$\deg\ is inferred   
(Eracleous \& Halpern 1998). While it is common to assume that in   
BLRGs the jet axis is parallel to the disk axis, so that $\theta \sim   
i$, with $i$ the disk inclination with respect to the line of sight,   
in this paper we will leave $i$ free to vary during the spectral fits   
to the X-ray data whenever possible.   
   
Optically, \3c\ is identified with a disturbed elliptical galaxy  
dominated by a very bright and unresolved nucleus (Matthews, Morgan \&  
Schmidt 1964; Martel et al. 1999), located in a moderately rich  
environment (Longair \& Seldner 1979). The optical spectra show a  
strong continuum and prominent broad lines photo-ionized by a power-law  
type of spectrum (Saunders et al. 1989; Tadhunter, Fosbury \& Quinn  
1989), with FWHM $\sim 12,000$ km/s for the H$\alpha$ line  
(Eracleous \& Halpern 1994). A recent estimate using the luminosity of  
the host galaxy places the mass of the central black hole around  
M$_{BH} \sim 10^9$ M$_{\sun}$ within 40\% (Marchesini et al. 2004).  
   
\3c\ is a variable source at X-ray (Dower et al. 1980; Barr \& Giommi  
1992), radio (Strom, Willis \& Willis 1978), optical, and UV  
frequencies (Puschell 1981; Tadhunter, Perez \& Fosbury 1986). At  
X-rays it is well studied, and was observed by all the major X-ray  
observatories before \suzaku. Flux and spectral variability is  
observed on time-scales shorter than a day, with a trend of spectral  
softening for increasing intensity (Gliozzi et al. 2007).   
 
Previous low-sensitivity observations of \3c\ indicated that its X-ray
spectrum was remarkably similar to Seyferts (Reynolds 1997). Aside
from the variable soft excess below 2~keV, \3c\ exhibits a warm
absorber with ionization parameter log$\xi \sim 2.5$ erg~cm~s$^{-1}$,
column density N$_H^W \sim 1-3 \times 10^{21}$ \nh, and outflow
velocity $\sim$1,000 km/s, clearly detected in recent high-resolution
dispersion spectra (Reeves et al. 2009; Torresi et al. 2010). The
X-ray continuum can be described by a power law with the ``canonical''
photon index $\Gamma \sim 1.8$ (e.g., Sambruna et al.~1999, S99
hereafter) and weak reflection features, including a Compton hump
above 10~keV with reflection fraction $R \sim 0.2-0.6$ (Gliozzi et al.~2007) 
and an
\feka\ line. The profile of the Fe~K line has remained, so far,
ambiguous due to its inherent complexity, as we show here (\S~5.1),
and the poor sensitivity of the pre-\suzaku\ detectors. For example,
using \asca\ an unusually broad (Gaussian width $\sigma_G$$\sim$1~keV)
line was inferred (S99; Reynolds 1997) while only a narrow component
with $\sigma_G$$\sim$100~eV was detected with \rxte, \sax, and the
\chandra\ HETGS (Eracleous et al. 2000; Grandi et al.~2001; Gliozzi et
al. 2007).
 
The source exhibits a strong, variable X-ray excess over the extrapolation  
of the power-law continuum from the hard X-rays below 2~keV. Extended  
soft X-ray emission was detected with the \rosat\ HRI in 0.3--2.4~keV  
(Prieto 2000), later confirmed with \chandra\ (Gliozzi et  
al. 2007). The contribution to the spectral soft excess from the  
extended emission, however, is negligible as shown by the point-like  
ACIS-S image in 0.5--2~keV (Gliozzi et al. 2007).    
  
Finally, while other BLRGs have been detected at GeV gamma-rays   
(3C~111; Hartman et al. 2008), no signal has been reported so far from   
\3c\ with \fermi\ (Abdo et al. 2010). The lack of gamma-ray flux and   
the double-lobe radio structure are strong indicators that the jet in   
this source is misaligned and most likely does not contaminate the   
nuclear emission, as supported by the Seyfert-like X-ray spectrum from   
the \asca\ and \rxte\ era.

\section{New X-ray Observations: Data Reduction}   
   
The log of the X-ray observations is reported in Table~1. The   
exposures are after data screening and the source count rates after   
background subtraction, according to the procedures described below.

\subsection{\suzaku\ XIS}    
   
\suzaku\ observed \3c\ on April 27th, 2007 for a net exposure time
after screening of $\sim$116~ks. We used the cleaned event files
obtained from version 2 of the \suzaku\ pipeline processing.  Standard
screening criteria were used, namely, only events outside the South
Atlantic Anomaly (SAA) as well as with an Earth elevation angle (ELV)
$ > 5\ensuremath{{}^{\circ }}$ were retained, and Earth day-time
elevation angles (DYE\_ELV) $ > 20\ensuremath
{{}^{\circ}}$. Furthermore, data within 256~s of the SAA passage were
excluded and a cut-off rigidity of $ >6 \,\mathrm{GV}$ was
adopted. The main parameters of the XIS observations are reported in
Table~1.
   
The XIS spectra of \3c\ were extracted from a circular region of
2.9$'$ radius centered on the source.  Background spectra were
extracted from four circular regions offset from the main target and
avoiding the serendipitous source and the calibration sources.  The
combined area of these four background regions is twice the area of
the main target region. The XIS response (rmf) and ancillary response
(arf) files were produced, using the latest calibration files
available, with the \textit{ftools} tasks \textit{xisrmfgen} and
\textit{xissimarfgen}, respectively.  The source spectra from the front illuminated (FI)
CCDs were summed, and fitted jointly with the back illuminated (BI), the XIS1, spectrum
after verifying that the data from the FI and BI CCDs were consistent
with each other. In fact, we checked that the 2--10~keV continuum
slope and fluxes in the two cases are consistent within the 2\%. 
The net XIS background-subtracted source spectra were binned with 
a minimum of 25 counts per energy bin to ensure that the $\chi^2$ goodness 
of fit can be applied to the spectral analysis. This binning is maintained
throughout the spectral analysis and only in some figures it was 
increased for clarity, when specified. Consistent results were found
by varying the binning in a range up to 100 counts/bin.

\subsection{\suzaku\ HXD}    
   
For the HXD-PIN data reduction and analysis we followed the latest
\suzaku\ data reduction guide (the ABC guide Version
2)\footnote[1]{http://heasarc.gsfc.nasa.gov/docs/suzaku/analysis/abc/}.
We used the rev2 data, which include all four cluster units, and the
best background available, which account for the instrumental
background (Kokubun et
al. 2007)\footnote[2]{ftp://legacy.gsfc.nasa.gov/suzaku/doc/hxd/suzakumemo-2008-03.pdf}.
The source and background spectra were extracted within the common
good time interval and the source spectrum was corrected for the
detector dead-time. The net exposure time after screening was 113~ks.
   
The contribution of the diffuse cosmic X-ray background counts was  
simulated using the spectral form of Boldt (1987), assuming the  
response matrix for diffuse emission, and then added to the  
instrumental one. Two instrumental background files are available;  
background A or ``quick'' background, and background D or ``tuned''  
background. We adopted the latter which is the latest release and  
which suffers lower systematic uncertainties of about 1.3\%,  
corresponding to about half the uncertainty of the first release of  
the Non X-ray Background.  With this choice of background, \3c\ is  
detected up to 70~keV with the PIN at a level of $\sim$20\% above the  
background. The count rate in 10--30~keV is 0.135 $\pm$ 0.003 c/s. The  
HXD PIN spectrum was binned in order to have a signal-to-noise ratio  
greater then 10 in each bin, and the latest response file released by  
the instrumental team was used.  
  
For a comparison with the \emph{Swift} BAT spectrum, we considered the  
PIN spectrum in the overlapping energy band of the two detectors,  
E$=$15--70~keV (see \S~3.4). In this band, the PIN spectrum can be  
fitted with a single power law with $\Gamma_{PIN}=1.95\pm0.14$,  
yielding a flux of $6.37^{+0.19}_{-1.23}\times 10^{-11}$ \flux, with  
90\% errors.

\subsection{\xmm}    
   
\xmm\ observed \3c\ on April 28th, 2008, for a net exposure after screening 
of 20~ks. The analysis of the RGS spectrum has already been   
performed by Torresi et al.~(2010) and is not reported here. 
   
We downloaded and reduced the data of the three EPIC cameras, pn, MOS1   
and MOS2. The data reduction was performed following the standard   
procedure with the XMM-SAS v. 8.0.1 package. We checked the   
observation for high background contamination, looking for flares in   
the light curves at energies greater than 10~keV. We excluded these   
bad intervals from the successive analysis, and extracted the source   
photons from a circular region of 40\arcsec, while the background ones   
were collected from an adjacent source free circular region of the   
same size. Only single and double events were selected.   
   
Using the SAS task \emph{epatplot} we checked that the pile-up  
fraction of the EPIC pn is negligible, less than 1\%. Instead, given  
the X-ray brightness of the source, the pile-up fraction of the EPIC  
MOS observations is significant, $\sim$50\%. The main result of  
pile-up is to introduce an overall distortion of the observed  
continuum, as it basically consists in counting two or more low-energy  
photons as a single high-energy photon. In fact, we find that the  
2--10~keV power-law continuum of the pn has a $\Gamma$$\simeq$1.74,  
while for the MOS $\Gamma$$\simeq$1.6. Therefore, given the  
significant pileup of the MOS and and the $\la$25\% lower S/N ratio of  
the MOS data, due to the short exposure and lower effective area,   
we limited the analysis of \xmm\ only to data from the EPIC pn  
camera.

Similarly to the XIS, the net EPIC pn background-subtracted source spectrum 
was binned with a minimum of 25 counts per energy bin to ensure that the $\chi^2$ 
goodness of fit can be applied to the spectral analysis
This binning is maintained throughout the successive spectral analysis and
increased only in some figures for clarity of presentation, when
specified. Consistent results were found by varying the binning in a
range up to 100 counts/bin.

\subsection{\swift\ BAT Observations}    
   
The BAT spectrum was derived from the 58-month hard X-ray
survey\footnote[1]{http://heasarc.gsfc.nasa.gov/docs/swift/results/bs58mon/}. The
data reduction and extraction procedure of the 8-channel spectrum is
described in Baumgartner et al. (2010).  To fit the BAT spectrum, we
used the latest calibration response \verb+diagonal_8.rsp+ and
background files as of December 2010.
  
The source is clearly detected in the energy band E$=$15--200~keV with  
a S/N$=$35 and a count rate $(1.35\pm0.05) \times 10^{-3}$ c/s.  A fit  
to the mean 58-month BAT spectrum in the 15--70~keV band with a single  
power law model yields a photon index $\Gamma_{BAT}=2.12\pm0.20$ and a  
flux of $5.10^{+0.16}_{-1.93} \times 10^{-11}$ \flux, at 90\%  
uncertainty. These values are consistent with those derived for the  
\suzaku\ HXD PIN in the same energy range (see \S3.2), justifying  
joint fits using both instruments. This is especially important in  
view of the extended sensitivity to higher energies provided by the  
BAT, which will allow us to set better constraints on reflection  
models.

\section{Timing Analysis}   
   
Figure~1 shows the \suzaku\ XIS0-3 light curves of \3c\ at soft and  
hard X-rays, respectively, with the flux in bins of 5760~sec (1  
orbit). The data from the three detectors are plotted on separate  
panels for comparison. Weak flux variations are seen in both energy bands for all 
three detectors, with small-amplitude variations of the order of few percent. 
When the ratio of the counts in the two energy bands is plotted against  
time or total count rate, there is no evidence for spectral variability.   
We note the presence of two possible short time-scale flares of
$\sim$5\% amplitude in the XIS~3. We ascribe them to spurious instrumental 
fluctuations. However, as discussed in the Appendix, they have a negligible 
influence on the spectral analysis of the average spectrum reported here.
Aside from the small apparent flares, the overall shape of the light curves of the 
different XIS detectors is consistent, with each showing a gradual increase in 
counts over the Suzaku observation.
  
The variability analysis was repeated for the \xmm\ EPIC data, despite  
the short exposure affected by the large background flares. No  
evidence for flux changes was found. Table~2 reports the soft  
(0.5--2~keV) and hard (2--10~keV) fluxes of \3c\ from the \suzaku\ and  
\xmm\ observations. The reported fluxes are both observed and corrected for 
any intervening absorption and were derived from the best-fit models to the  
0.6--200~keV \suzaku+BAT and 0.5--10~keV EPIC pn spectra  
(see \S~5.4 and \S~6). Observed and absorption corrected luminosities are
reported in Table~2 as well. Comparing both fluxes and luminosities in  
all energy bands, it is clear that the source was in similar intensity  
states at the epoch of the \suzaku\ and \xmm\ observations.

\section{The \suzaku\ Broad-Band Spectrum}   
  
The lack of significant spectral variability derived from a
model-independent timing analysis of the XIS data (\S~4) indicates 
that the spectral
properties of \3c\ can be safely investigated using a time-averaged
spectrum accumulated over the entire 116~ks exposure. This spectrum
provides the largest signal-to-noise ratio available from this
observation, allowing the most detailed spectral analysis of the Fe~K
line region and broad-band continuum, the main goals of our project.
  
All spectral fits were performed using \verb+XSPEC+ v.12.6.0. The  
significance of adding free parameters to the model was evaluated with  
the F-test. 
We included only model components with a statistical probability $P_F$$>$99\%. 
All uncertainties  
quoted are 1$\sigma$ for one parameter of interest, unless otherwise  
stated. The energy of the lines is reported in the source rest-frame,  
if not otherwise stated. The data from the XIS CCDs were fitted  
jointly, excluding the energy range E$=$1.5--2~keV, around the Si K edge,  
which is known to be affected by calibration issues (Koyama et al.~2007).  
When fitting the broad-band spectrum, a constant factor was  
introduced in the model to take into account the cross-normalization  
between the different instruments. In particular, a constant factor  
of $\simeq$1.16 was adopted for the XIS/PIN and of $\simeq$0.9 for the 
PIN/BAT. Throughout the spectral analysis we always include a Galactic 
absorption towards \3c\ of $N_H$$=$$7.4\times10^{20}$ \nh\ 
(e.g., Dickey \& Lockman 1990; Kalberla et al.~2005). 
  
The plan of the analysis is as follows. First, we consider the combined 
broad-band \suzaku\ and \swift\ BAT spectra in the E$=$0.6--200~keV, 
excluding the Fe~K band, and derive an estimate of the underlying X-ray 
continuum. Then, we consider also the Fe~K band and perform a search for 
additional emission/absorption features in the spectrum. Subsequently, we perform 
fits to the full broad-band spectrum with a physical model including all the 
spectral components identified in the previous steps plus an additional 
relativistic line and smeared ionized reflection from the putative accretion
disk. Finally, we test the simultaneous alternative modeling of the soft excess, Fe K lines
and cold reflection with a second lowly ionized reflection component.

\subsection{The Continuum Determination}

The combination of the \suzaku\ and \swift\ BAT instruments provides
an unprecedentedly wide and sensitive spectral coverage from 0.6~keV
up to 200~keV. This is fundamental for a correct determination of the
underlying X-ray continuum. From this analysis, we exclude the Fe~K energy
range, E$=$5.5--7.5~keV, because, as discussed in the next section,
the presence of spectral complexities in this interval can affect the
best-fit continuum model. 

In the upper panel of Figure~2 we show the ratio of the combined XIS03,
XIS1, PIN, and BAT spectra with respect to a Galactic absorbed
power-law continuum. As seen, this simple model does not provide
a sufficient parameterization of the data ($\chi^2/dof$$=$5027/3166)
and strong deviations are present at both low and high energies. The
resultant power-law photon index is $\Gamma$$\sim$1.88.

At energies lower than $\sim$2~keV there is evidence for a soft excess
component. This is expected also from previous studies of \3c, see
\S~2. Following previous authors, we initially model the soft excess with a 
phenomenological blackbody with best-fit temperature $kT$$\simeq$100~eV. 
This is highly required,
with a $\Delta\chi^2$$\simeq$1407 for two additional model
parameters. The alternative phenomenological
modeling with a power-law provides a comparable fit. However, the
physical interpretation of the resultant very steep photon index of
$\Gamma$$\sim$5 is more ambiguous and therefore in the following we
keep the blackbody parameterization.

The excess flux in the upper panel of Figure~2 between 10--40~keV is
instead indicative of a reflection component. This is also required on
a physical basis by the presence of a narrow \feka\ line
(e.g., Lightman et al. 1988), as discussed in the next section.
Therefore, we added a cold reflection component using \emph{pexrav} in
\verb+XSPEC+. We assume an inclination angle of $i$$=$30$^\circ$,
typical for BLRGs, and standard solar abundances.  This component is
highly required, with a $\Delta\chi^2$$\simeq$66 for two model
parameters, significant at $>$99.99\%.  The power-law continuum slope
is now $\Gamma$$\simeq$1.85, the high energy cut-off
E$_c$$\sim$200--300~keV, and the reflection fraction from \emph{pexrav} is 
R$\simeq$0.36. The final fit is good, with
a reduced chi-squared very close to unity, $\chi^2/dof$$=$3554/3162.  
The ratios of the spectra with respect to this best-fit continuum model, including 
the Fe K band, are shown in the middle panel of Figure~2.

In conclusion, the broad-band X-ray continuum of \3c\ is well
parameterized by a Galactic absorbed power-law, plus a blackbody soft
excess and a reflection component. We use this as the baseline
continuum model for the successive search of additional
emission/absorption spectral features.

\subsection{Search for Emission/Absorption Spectral Features}

Looking at the spectral ratios in the middle panel of Figure~2, weak
dips indicating absorption can be seen at E$<$2~keV, as
well as further spectral complexities in the Fe~K region.  The BLRG
\3c\ is known to exhibit absorption features at energies below 2~keV
due to a warm absorber (Reynolds 1997), recently studied with the
\chandra\ and \xmm\ gratings (Reeves et al. 2009; Torresi et
al. 2010). Despite the XIS lower resolution compared to the gratings,
absorption features are visible in Figure~2 (middle panel) around
0.8--1~keV, possibly due to ionized O and Fe~L. We thus include a warm
absorber component in our fit, modeled by an \verb+XSTAR+ grid
(Kallman \& Bautista 2001).  Given the limited S/N and energy
resolution available with respect to the gratings studies, we fix the
gas outflow velocity to 1,000 km/s, consistent with what found by
Reeves et al.~(2009) and Torresi et al.~(2010). We assumed an ionizing
X-ray power-law with photon index $\Gamma$$=$2 and turbulence velocity
of 500~km/s, similar to what assumed by Reeves et al.~(2009).

The inclusion of the warm absorber component is highly significant,
$\Delta\chi^2$$=$30 for two additional parameters, corresponding to a
detection probability $\ga$99.99\%. We estimate a column density
N$_H$$\simeq$$6\times 10^{20}$~cm$^{-2}$ and an ionization parameter
log$\xi$$\simeq$$2.44$~erg~s$^{-1}$~cm.

We now reintroduce the E$=$5.5--7.5~keV energy band in the fits. Up to
now the fit statistics is $\chi^2/dof = 3524/3160$. Including the Fe~K
band, this changes to $\chi^2/dof = 4760/4177$. The middle panel of
Figure~2 and upper panel of Figure~3 show the presence of spectral
complexities in the Fe~K band. In particular, a prominent emission
line is visible at rest-frame energy $\sim$6.4~keV, together with
several other emission lines in the energy range $\sim$5.5--7.5~keV.

As a first step, we modeled these emission features with
Gaussians. The most prominent line at E$=6.41\pm0.01$~keV is clearly
identified with the \feka\ from lowly ionized, fluorescent material. The line
is resolved, with a width of $\sigma_G$$=$$118^{+20}_{-13}$~eV,
equivalent width EW$=$$68\pm9$~eV, and flux I$=$$(3.3\pm0.2)\times
10^{-5}$~ph~s$^{-1}$~cm$^{-2}$. The inclusion of this feature yields a
$\Delta\chi^2$$=$233 for three additional model parameters, which
corresponds to a detection probability $\gg$99.99\%.

Next, we added the associated \fekb\ component. In the fit, the energy
was fixed to the expected value of E$=$7.06~keV, the width was set
equal to that of the \feka\ and its intensity to $\simeq$13\% of the
\feka\ (e.g., Molendi, Bianchi \& Matt 2003 and references therein).
The addition of the \fekb\ improves the fit by $\Delta\chi^2$=13.

Three additional emission features are still present in the spectral  
residuals at energies of E$\sim$5.8, 6.9, and 7.5~keV. To account for  
these, we added three more Gaussians to the model.  The two lines at  
higher energies are not resolved. Their energies are at  
E$=$$6.91\pm0.02$~keV and E$=$$7.51\pm0.03$~keV and have equivalent  
widths of $17\pm6$~eV and $18\pm7$~eV, respectively. The  
$\Delta\chi^2$ is 21 and 19, for two additional model parameters,  
which correspond to detection probabilities of $\simeq$99.99\%.  The lower  
energy line at E$=$$5.70\pm0.09$~keV is instead broader and resolved,  
with a width of $\sigma$$=385^{+77}_{-57}$~eV and equivalent width of  
$55\pm 12$~eV. Its significance is higher, $\Delta\chi^2$$=$58  
for three additional parameters, corresponding to a detection   
probability $>$99.99\%. 

The fit with the five Gaussian lines is purely phenomenological, in
order to have a first representation of the Fe~K complex and a
determination of the lines parameters and their significance. All the
Gaussian lines are detected with a probability $\ga$99.99\%.
Concerning the identification of the three additional emission lines,
the feature at 7.5~keV may well be identified with emission from lowly 
ionized Ni~K$\alpha$, whose atomic transition energy
is expected at E$\simeq$7.5~keV (e.g., Molendi, Bianchi \& Matt 2003
and references therein).  The two remaining lines at 6.9~keV and
5.8~keV are at energies not directly consistent with any significant
atomic transitions and therefore have a less obvious physical
interpretation.  The closest atomic lines to the 6.9~keV line are the
Fe~XXV He$\alpha$ (1s$^2$--1s2p) at E$\simeq$6.7~keV and Fe~XXVI
Ly$\alpha$ (1s--2p) at E$\simeq$6.97~keV (Kallman et
al.~2004). However, the identification with one of these lines would
require a blue/red-shift of $\simeq$0.03c or $\simeq$0.01c,
respectively. The 5.8~keV line is even more mysterious, being out of
the range expected for Fe~K lines, from 6.4~keV up to 7~keV, depending
on the ionization. If identified with Fe~K emission, the large width
and red-shift could suggest strong relativistic effects and an origin
close to the central black hole (see next section). 

The inclusion of the warm absorber and the emission lines in the Fe K
band on the continuum model provides a very good fit, with a
$\chi^2/dof$$=$4416/4167. Therefore, the final best-fit model of the
combined broad-band \suzaku\ and \swift\ BAT spectrum of \3c\ is
rather complex and is composed by a Galactic absorber power-law
continuum, a black body soft excess, a cold reflection component, a
warm absorber and five Gaussian emission lines in the Fe K band. The
best-fit parameters are reported in Table~3. The spectral ratios with
respect to this model are shown in the lower panel of Fiure.~2.

\subsection{A Relativistic Fe~K Emission Line in 3C~382?}  
   
Inspection of Figure~3 (upper panel) suggests the intriguing possibility 
that the 5.7~keV and 6.9~keV emission features represent two components 
of a single, broader feature, i.e. a relativistic Fe~K line. More 
precisely, they would correspond to the broad red-wing and sharp 
blue-peak components of such a line, respectively (e.g., Fabian et 
al. 1989). As an initial test to this hypothesis we replaced these 
two features with a single relativistic line component, \emph{diskline} 
in \verb+XSPEC+ (Fabian et al.~1989), in the model discussed in the 
previous section and reported in Table~3. Therefore, now the Fe~K band is
composed by three narrow emission lines, an \feka\ and 
associated Fe K$\beta$ and Ni K$\alpha$, and a broad relativistic line 
(see middle panel of Figure~3). 

The \emph{diskline} model considers a non-rotating Schwarzschild black hole, 
with spin parameter $a$$=$0. Given that the fit is not sensitive to the 
power-law emissivity profile, we assumed the typical value of 
$\beta$$\simeq$$-2.5$ (e.g., Nandra et al.~2007; de la Calle Perez 
et al.~2010; Patrick et al.~2010). However, we checked that a 
different choice of $\beta$ in the wide range from $-5$ to 0 does 
not significanly change the parameters estimates reported below, 
being always consistent at the 90\% level. 

Interestingly, if the energy of the line is left free to vary, the fit
yields a value larger than the expected 6.4~keV for neutral \feka\ emission. 
The formal best-fit centroid is at E$=$6.69~keV,
ranging between 6.55 and 7.10~keV at 90\% confidence. This indicates
that the relativistic Fe~K emission originates from ionized material,
preferentially Fe~XXV He$\alpha$ at 6.7~keV rest-frame, but we can not
clearly exclude also a contribution from Fe~XXVI Ly$\alpha$ at
6.97~keV. This supports earlier models claiming that the accretion
disk in BLRGs is ionized (Ballantyne et al. 2002), and adds to
previous evidence from the \suzaku\ observations of two other
classical BLRGs, 3C~390.3 (Sambruna et al. 2009) and 3C~120 (Kataoka
et al. 2007). 
 
Thus, we performed fits assuming, in turn, either emission from Fe~XXV  
or Fe~XXVI. In both cases the fit is highly improved, $\Delta\chi^2$$=$60  
for 4 additional parameters, corresponding to a detection significance  
$>$99.99\%.  The middle panel of Figure~3 illustrates  
the best-fit model for the three Gaussians and the \emph{diskline}  
profile. Assuming Fe~XXV (Fe~XXVI) emission, we obtain good constraints on  
the disk inclination $i=30^\circ \pm 1^\circ$ ($26^\circ \pm 2^\circ$), on  
the inner and outer radii of the emitting region, $r_{in} =12 \pm 2$~$r_g$  
($10\pm2$~$r_g$) and $r_{out}=23 \pm 3$~$r_g$ ($18\pm3$~$r_g$). 
Indeed, a tight estimate of $i$ was expected from the 6.9~keV peak  
sharp fall to the blue (see Figure~3, middle panel), while the limited  
extension of the red ``wing'' indicates that the inner radius can not  
extend too further in close to the black hole. The equivalent width of  
the line is EW$=$$80\pm25$~eV and the intensity I$=$$(3.2\pm0.4)\times  
10^{-5}$~ph~s$^{-1}$~cm$^{-2}$. 

We stress that the inclusion of the warm absorber has a negligible
effect on the Fe~K parameters (e.g., Reeves et al.~2004 and references
therein). This can be explained by the low level of ionization and
small column density of the warm absorber, which implies that only
the light elements (C, O, Mg, Ne) are contributing to the absorption
at energies below 2~keV. 
 
The final fit with the relativistic line profile is very good,
$\chi^2/dof=$4435/4168 and the data/model ratio with respect to this
model are shown in the lower panel of Figure~3.  Given the high quality
of the XIS data in 4--10~keV and good determination of the underlying
continuum in the wide 0.6--200~keV band, we regard the detection of a
broad, relativistic Fe~K line in \3c\ as robust. In support of this,
we performed extensive calibration and background tests, which
are described in the Appendix.

\subsection{Detailed Modeling of the Relativistic Line and Ionized Reflection} 

Here we discuss in more details the modeling of the broad Fe K line and 
possible ionized reflection component in the broad-band 0.6--200~keV spectrum.
We first examine the broad \feka\ profile in light of the possibility 
that it could be due to a rotating (either prograde or retrograde) 
black hole. This is possible by using the relativistic line code 
\emph{relline} of Dauser et al. (2010), which allows also the black 
hole spin $a$ to vary between $-$0.998 and $+$0.998. We use the final 
best-fit model discussed in \S~5.2 and reported in Table~3, substituting 
the 5.7~keV and 6.9~keV emission lines with a single broad relativistic line. 

The main parameters of \emph{relline} are the rest-frame energy of the
line, E, the power-law emissivity profile, $\beta$, the black hole
spin, $a$, the inner and outer radii of the reflecting region,
$r_{in}$ and $r_{out}$, the disk inclination, $i$, and the intensity
of the emission line, $I$.  At first, all the above parameters were
free to vary during the fit. Only $r_{in}$, $r_{out}$, $i$, $I$ were
constrained. We explored the parameter space in more detail, checking
the effects of assuming different values. However, given the limited
S/N of the present data and the complexity of the model, we are unable
to constrain the spin, emissivity and energy all together.
Nevertheless, good fits are obtained for $|\beta|$$\simeq$2--3.  The
line energy is E$\sim$6.6--7~keV at 90\%, confirming ionized
emission. Similar fits are obtained assuming Fe XXV~He$\alpha$ at
E$=$6.7~keV or Fe~XXVI Ly$\alpha$ at E$=$6.97~keV, with slightly
smaller $r_{in}$ and $r_{out}$ in the latter case.

We also explored the dependence of the line shape on the black hole
spin by imposing that the inner radius coincide with the Innermost
Stable Circular Orbit (ISCO) of the disk. This corresponds to assuming that 
the beginning of the reflection region is coincident with the inner radius of 
the disk and that the latter extends down to the ISCO. The ISCO assumes the
following values depending on $a$: $r_{ISCO}$$\simeq$1.23$r_g$, 6$r_g$ 
and 9$r_g$ for $a$$=$$+0.998$, $0$ and $-0.998$, respectively. A series of 
fits were performed with the inner radius linked to the ISCO parameter in the 
\verb+relline+ model in \verb+XSPEC+, and leaving $a$ free to vary. However, 
also in this case we can not still constrain all the parameters simultaneously 
and find a clear deep minimum in the statistical distribution. For instance, 
given the values in the intervals $\beta$$=$1--3, E$=$6.7--6.97~keV and
$a$$=$$-0.998$--$+0.998$, there are several minima, all within a 
difference of a few in $\Delta\chi^2$, and the fit does not reach a 
stable convergence. 

Therefore, despite the good quality of the present \suzaku\ data  
and the broad-band coverage up to 200~keV, no sensitive constraint was  
possible on the value of the black hole spin, i.e., we obtain similar  
fits for $a=-0.998$, $a=0$, and $a=+0.998$. Detailed simulations show  
that much larger exposures are needed, $\sim$ 300~ks, to achieve  
the goal of measuring $a$ with sufficient accuracy to distinguish  
among a static, prograde or retrograde spinning hole. 

Given the results described so far, in the remaining fits to the
0.6--200~keV spectrum of \3c\ the broad Fe~K line energy was fixed to
E$=$6.7~keV, $a$$=$0, and $\beta=-2.5$, an intermediate value
consistent with Seyferts (Nandra et al. 1997, Patrick et al. 2010, de
la Calle Perez et al.~2010).  Including this component, the fit is
improved by $\Delta\chi^2$$=$60 for 4 additional parameters,
corresponding to a high detection probability $>$99.99\%. This fit
yields $r_{in}$$=$$12\pm2$~$r_g$ and $r_{out}$$=$$23^{+4}_{-2}$~$r_g$,
with $i$$=$$30^{\circ}\pm 1^{\circ}$. The line intensity is
$I$$=$$(3.0\pm0.4)\times 10^{-5}$~ph~s$^{-1}$~cm$^{-2}$ and
EW$=$$76\pm18$~eV. This indicates that the broad, ionized Fe~K line is
produced from a narrow annulus of $r_{in}$$\sim$10$r_g$ and
$r_{out}$$\sim$20$r_g$, i.e., at some distance from the black hole. We
checked that a different choice of the $\beta$ in the range $-2$ to
$-3$ does not change significantly the parameters estimates, being
always consistent within the 90\% errors.

As discussed above, the presence of ionized Fe K emission lines in the
energy range 6--7~keV provides strong motivation for including an
additional ionized reflection component to our baseline continuum
model.  Sophisticated models are known in the literature that include
the proper number of emission lines expected for the fitted ionization
parameter, $\xi$ (e.g., Ross, Fabian \& Young 1999; Ross \& Fabian
2005; Garc{\'{\i}}a \& Kallman 2010).  Two such models,
\emph{reflionx} (Ross \& Fabian 2005) and the more recent
\emph{xillver} (Garc{\'{\i}}a \& Kallman 2010), are available in
\verb+XSPEC+ for detailed fits. We choose to use the latter, convolved
with the \emph{relconv} model of Dauser et al.~(2010), which takes
into account line blurring due to relativistic effects.  We use a
\emph{xillver} table with log$\xi$$=$0.8-3.8~erg~s$^{-1}$~cm, standard
solar abundances, and an incident ionizing power-law with
$\Gamma$$=$2.  Consistent results were obtained using \emph{reflionx}.
 
The free parameters of the \emph{relconv} component are the power-law
emissivity profile, inner/outer radii on the disk surface, spin, and
inclination angle. The free parameters of \emph{xillver} are instead
the ionization level of the material and the normalization. During the
first fits to the 0.6--200~keV spectrum, these parameters were left
free to vary. The fits yielded $r_{in}$$\simeq$10~$r_g$,
$r_{out}$$\simeq$25~$r_g$, inclination $i$$\simeq$30$^{\circ}$, and
ionization parameter log$\xi$$\simeq$3~erg~s$^{-1}$~cm, while the disk
emissivity $\beta$ was less constrained and the spin $a$ was
unconstrained.  After performing a series of tests, we fixed
$\beta$$=$$-2.5$, as no significant changes in the line and continuum
parameters were observed within 90\% uncertainties if the emissivity
profile was left free to vary in the most probable range
$|\beta|$$=$2--3.  We then linked $r_{in}$$=$$r_{ISCO}$ in
\emph{relconv}, and performed three separate fits for $a=+0.998$, $0$,
and $-0.998$, to explore the dependence on the spin. A slightly better
fit was obtained for the case of negative spin, with
$\Delta\chi^2$$\simeq$2 and $\simeq$5 higher with respect to the cases
with zero or positive rotation. Thus, we conclude again that the
quality of the current \suzaku\ data of \3c\ does not allow us to
constrain the sign of the black hole spin. Future deeper observations
($\ga$300~ks) are needed to this end. However, given that the fit is
slightly better for a spin of $-0.998$ and that the overall model
parameters are still consistent at the 90\% whatever spin is assumed,
in the following fits and in Table~4 we fix the spin to this
value. The inclusion of the ionized reflection component improves the
fit by $\Delta\chi^2=65$ for four additional model parameters, which
correspond to a detection probability of $\gg$99.99\%.

The estimated ionization of the disk reflecting material is
log$\xi=3.04^{+0.03}_{-0.04}$ erg~s$^{-1}$~cm, consistent with the Fe
ion population being dominated by Fe~XXV, with the He$\alpha$ at
E$\simeq$6.7~keV being the most intense associated resonant line
(e.g., Ross, Fabian \& Young 1999; Ross \& Fabian 2005; Garc{\'{\i}}a
\& Kallman 2010). The fitted outer radius is
$r_{out}$$=$$26^{+6}_{-4}$~$r_g$ and the inclination
$i$$=$$29^{\circ}\pm2^{\circ}$, consistent with the more simplistic
fits using \emph{diskline} or \emph{relline}. The fit is very good, 
 with a reduced chi-squared very close to unity, $\chi^2/dof = 4430/4168$.
The best-fit parameters of this broad-band fit are
reported in Table~4, while the data, best-fit model and spectral
ratios are shown in the left panels of Figure~4.

\subsection{Two ionized reflection components in 3C~382?}

It should be noted that the modeling of the soft excess with a simple black body component in the previous sections, besides providing a good representation of the data, might be difficult to explain from a physical point of view. 
In particular, from a systematic X-ray spectral analysis of PG quasars, Gierlinski \& Done (2004) demonstrated that if the soft excess is modeled with a simple black body component, the resultant temperature is in the narrow range $kT$$\sim$100--200~eV, despite the large range in Eddington luminosities and black hole masses of the sources. The temperature is also too high to be directly associated with standard Shakura-Sunyaev accretion disks.
This suggests a possible atomic origin for the soft excess and a promising alternative physical explanation is in terms of a blurred ionized reflection component (e.g., Crummy et al.~2006). However, the modeling of the soft excess is still required after the inclusion of the ionized disk reflection component in the previous section. When parameterized with a black body, it still provides a high improvement of the fit, $\Delta\chi^2$$=$538 for two model parameters, which corresponds to a detection probability $\gg$99.99\%.

The presence of an additional mildly ionized reflection component is required to self consistently model the Fe K$\alpha$/K$\beta$, Ni K$\alpha$ emission lines and reflection hump. 
In particular, the energy of the Fe K$\alpha$ is $\sim$6.4~keV for a wide range of ionization, from Fe~I to Fe~XVII (Kallman et al.~2004). This gives rise to the possibility of simultaneously modeling the soft excess, Fe K lines and reflection hump with an additional reflection component. Therefore, we performed a test replacing the previous \emph{pexrav}, black body and Fe K Gaussian emission lines with a cutoff power-law continuum and an additional \emph{xillver} reflection component. To account for the velocity broadening, we convolved the latter with a Gaussian profile with the same width of the Fe K$\alpha$ emission line, i.e., $\sigma$$=$90~eV, which corresponds to FWHM$\sim$10,000~km/s (see Table~4).

The best-fit values are reported in Table~5. The power-law continuum, warm absorber and relativistic highly ionized reflection parameters are essentially unchanged. The ionization parameter of the mildly ionized reflection is log$\xi$$=$$1.5\pm0.03$~erg~s$^{-1}$~cm. This is consistent with Fe~XIII--XVII being the most abundant Fe ions (Kallman et al.~2004; Garc{\'{\i}}a, Kallman \& Mushotzky 2011). As shown in the right panels of Figure~4, this provides a very good fit, with $\chi^2$$=$4467 for 4174 degrees of freedom. The fit is only slightly worse than that with the model in Table~4. The reduced $\chi^2$ is 1.06 and 1.07 using the previous and present model, respectively. However, the present model has the strong advantage to physically self-consistently explain all the main spectral features in the broad-band spectrum with the lowest number of model components. This result is remarkably similar to what observed for local Seyfert galaxies (e.g., Nandra et al.~2007; de la Calle Perez et al.~2010; Patrick et al.~2010).

The reflection fraction $R$ is conventionally used to estimate the relative contribution of the reflection component with respect to the power-law continuum emission. This parameter is directly derived when using the \emph{pexrav} model in \verb+XSPEC+, as in Table~3 and Table~4. This is related to the solid angle covered by the reflecting gas as $R$$=$$\Omega/2\pi$. However, anisotropies, obscuration and relativistic effects close to the black hole can all influence this simple conventional relation for AGN, to the point that it can become physically meaningless (Crummy et al.~2006; Garc{\'{\i}}a, Kallman \& Mushotzky 2011). Therefore, we define a phenomenological \emph{reflection flux fraction}, $R_F$, as the ratio between the unabsorbed 0.5--100~keV flux of the reflection component and continuum. Its value is still comparable to the conventional parameter, $R_F$$\sim$$R$ (Garc{\'{\i}}a, Kallman \& Mushotzky 2011).
As reported in Table~5, this fraction is about 0.1 for both highly and mildly ionized reflection components, which means that each of them contributes $\sim$10\% on the overall broad-band spectrum. In particular, the mildly (highly) ionized component contributes $\sim$20\% ($\sim$30\%) in the 0.5--2~keV, $\sim$5\% ($\sim$10\%) in the 2--10~keV and $\sim$10\% ($\sim$5\%) in the 10--200~keV with respect to the power-law continuum (see Figure~4 right, upper panel). 

Due to the wide-band coverage allowed by the BAT, very good constraints can be obtained on the power-law continuum energy cutoff. The latter value of $E_c$$\simeq$180~keV is completely consistent for both models presented in this section and \S~5.4, see Table~4 and Table~5. This clearly suggests a thermal disk-corona origin for the primary power-law. Thus, it is unlikely that the X-ray power-law continuum of \3c is due to a jet. 

In conclusion, the broad-band 0.6--200~keV spectrum of \3c\ can be described by a
very complex model composed by a Galactic absorbed power-law
continuum, a black body soft excess, a warm absorber, an Fe K$\alpha$
emission line and related Fe K$\beta$, a Ni K$\alpha$ emission line,
a mildly ionized reflection component and a relativistic smeared ionized
reflection component. Alternatively, it can be well represented also by a Galactic absorbed cutoff power-law continuum, a warm absorber and two ionized reflection components: a mildly ionized one which simultaneously reproduces the narrow Fe K$\alpha$ emission line, soft excess and high energy hump and a highly ionized one which reproduces the broad relativistic Fe K line.
These two models are indistinguishable from a statistical point of view, but the latter provides a more physically self-consistent representation of the data, with the lowest number of model components.

\section{The \xmm\ Spectrum}    
  
The spectral analysis of the EPIC pn observations was carried out   
using the \emph{heasoft} v.~6.5.1 package. 
As shown in Figure~5, the overall 0.6-10~keV pn spectrum is very
similar to the XIS one, albeit at lower S/N due to the lower exposure
(factor 6) which compensates for the larger EPIC effective area
compared to the XIS cameras. The ratios against a Galactic absorbed
power-law continuum are reported in the middle panel of Figure~5. As we
can see, the fit is very poor ($\chi^2/dof = 4149/1454$) and the
power-law photon index is $\Gamma \simeq 2$. Intense emission
residuals are present at low and high energes, similar to the
\suzaku\ ones in the upper panel of Figure~2. These deviations indicate
the presence of a soft excess and a reflection components. 

Because of the lower quality of the data and the spectral
similarities, and because the source was in similar intensity states
at both epochs (Table~2), we assumed for the EPIC pn the best-fit
model previously derived for the \suzaku\ data from the fits to the
broad-band spectrum discussed in \S~5.2 and reported in Table~3. We
assumed only a cold reflection component modeled with \emph{pexrav} and 
fixed the cut-off energy and reflection fraction to their best-fit values. 
We also assumed only the presence of the narrow Fe K$\alpha$/K$\beta$ emission 
lines, as indicated by the data.

The fit to the EPIC pn data with this model is good,
$\chi^2/dof$$=$1567/1448. The best-fit model and spectral ratios are
reported in the upper and lower panels of Figure~5, respectively.  The
power-law continuum photon index is $\Gamma$$=$$1.87\pm0.01$. The EPIC
data confirm the presence of a strong soft excess, with a blackbody
temperature k$T$$=$$142\pm5$~eV and normalization $(1.72\pm0.03)\times
10^{-4}$. The warm absorber is also detected, with column density
$N_H$$=$$1.8^{+0.5}_{-0.2}\times 10^{21}$~cm$^{-2}$ and ionization
parameter log$\xi$$=$$2.35\pm0.05$~erg~s$^{-1}$~cm, consistent with
\suzaku\ (\S~5.2). 

  
An unresolved ($\sigma$$=$100~eV) \feka\ line is detected at  
E$=$$6.4\pm0.03$~keV, with a flux $I$$=$$(2.5\pm0.4)\times  
10^{-5}$~ph~s$^{-1}$~cm$^{-2}$ and EW$=$$54^{+22}_{-14}$~eV.  Due to  
the lower S/N of the pn data, there is no clear evidence for the  
associated \fekb\ and Ni lines present in the XIS data, or for a  
broad, relativistic Fe~K line at 6.67~keV. The 90\% upper limit for  
the latter is EW$<$130~eV.    
  
In summary, an archival 20~ks \xmm\ EPIC pn observation of \3c\ was  
analyzed and found consistent with the best-fit model to the  
\suzaku\ XIS data. The observed and intrinsic (corrected from any 
intervening absorption) fluxes and luminosities in the 0.5--2~keV and 
2--10~keV bands from the EPIC pn data are reported in Table~2.

\section{Summary of Observational Results}    
   
This paper presented a 116~ks \suzaku\ observation of the Broad-Line  
Radio Galaxy \3c\ ($z$=0.057). A shorter (20~ks), unpublished  
\xmm\ EPIC exposure was discussed, as well as a \swift\ BAT spectrum  
from integrating 5-years of survey observations. The main results of  
the data analysis are:  

\noindent{\it The soft X-ray spectrum - } Below 2~keV, the XIS data   
indicate the presence of a warm absorber, with parameters consistent   
with those derived from the analysis of a previous \chandra\ HETGS   
spectrum. The soft excess previously observed is confirmed by both   
\suzaku\ and \xmm\, with an increase in flux between the two epochs of 
 $\sim$10\%. This component can be well modeled by a phenomenological
black body with $kT$$\sim$100~keV, but a more physical parameterization 
suggests a possible origin from mildly ionized reflecting material.

\noindent{\it The Fe~K Region - } An intense narrow Fe K$\alpha$ 
emission line is detected at E$\simeq$6.4~keV, together with the associated 
Fe K$\beta$ component. An additional weak unresolved emission line at 
E$\simeq$7.5~keV is tentatively observed, possibly identifiable with
Ni K$\alpha$. However, more interestingly, a prominent Fe~K line with a  
relativistic profile is found; its center energy is consistent at 90\%  
confidence with emission from ionized Fe, either or both FeXXV/FeXXVI  
in the energy range 6.55--7.10~keV. The profile is consistent with  
emission from a disk annulus between inner and outer radii $r_{in}  
\sim 10$~$r_g$ and $r_{out} \sim 20$~$r_g$, respectively, for an  
assumed disk emissivity $\beta=-2.5$. An ionized disk reflection model
was included to parameterize this component in the broad-band spectrum.
   
\noindent{\it The spectrum at energies $\ga$10~keV - } The data at   
medium and hard X-rays are well described by a power law with   
$\Gamma\simeq1.8$, cutting off around E$_c\sim 200$~keV. Above 10~keV,  
contributions from both mildly and highly ionized reflection are present.
Each of them contribute $\sim$10\% on the overall broad-band continuum,
but the mildly ionized one has an higher impact at E$>$10~keV.

\section{Discussion}    
   
This paper presents new results from the analysis of our \suzaku\, as 
well as archival \xmm\ and \swift\ BAT observations of \3c, one of 
the ``classical'', best-studied BLRGs from optical and radio 
samples. The combination of the \suzaku\ and BAT data provides an 
unprecedented sensitive coverage in the energy range 0.6--200~keV, 
useful to disentangle the various contributions to the total nuclear 
X-ray emission.  At first sight, the broad-band X-ray spectrum of 
\3c\ is remarkably Seyfert-like with most, if not all, the 
characteristics of a typical nearby radio-quiet AGN: a $\Gamma \sim 
1.8$ power-law continuum, a warm absorber, soft excess, relativistic 
\feka\ line, and a reflection bump above 10~keV. While typically 
radio-loud in other wavelengths, with giant radio lobes and a 
radio-to-optical Spectral Energy Distribution typical of other BLRGs 
(e.g., Grandi et al.~2001), \3c\ appears exclusively radio-quiet from 
an X-ray spectroscopy perspective. 
 
The intrinsic photon index we measure for \3c\ after accounting for 
the reflection components is $\Gamma$$\sim$1.8, similar to most radio-quiet  
Seyfert 1s observed with \suzaku\ (e.g., Patrick et  
al. 2010). Previous observations of \3c\ at X-rays (\S~2) established  
that the photon index varies in unison with the flux, in the sense of  
a softer spectrum for increasing X-ray intensity (Gliozzi et al. 2007  
and references therein), a trend \3c\ shares with radio-quiet,  
non-jetted AGN.  Together with the constraints on the high-energy  
cutoff of the primary power law, this implies that thermal  
Comptonization dominates the emission below 100~keV in \3c, supporting  
the idea that the bulk of the X-ray continuum does not originate in a  
jet. At present, only an upper limit is reported to the gamma-ray GeV  
emission at the location of \3c\ in the Fermi 15-months catalog (Abdo  
et al.~2010). This suggests that the contribution of  
a jet to the higher energies is weak, or occurs with a small duty  
cycle. Future \fermi\ monitoring will be necessary to shed light on the  
jet activity at high energies from the core of \3c.    
 
The high quality and low background of the XIS data allows us to model 
with high fidelity the Fe-K emission. We confirm the presence of a 
narrow line at an energy of $\sim$6.4~keV, identified with the narrow component 
of the \feka\ line. 
A similar narrow component, consistent with 
emission from the Broad Line Region (BLR), was detected in 3C~390.3 (S09) and 3C~120 
(Kataoka et al. 2007), two of a handful of well-known, nearby BLRGs 
observed with \suzaku\, and is regularly observed in Seyfert 1s 
(Patrick et al. 2010; Nandra et al. 2007).
This line can be modeled with a mildly ionized reflection component with 
log$\xi$$\simeq$ 1.5~erg~s$^{-1}$~cm (Table~5), which can also simultaneously
explain the observed soft excess and part of the reflection hump. This 
contributes $\sim$10\% on the power-law continuum.

From the width of the line of FWHM$\sim$10,000~km/s and  assuming that the 
material is in Keplerian motion in the gravitational potential well of the 
super-massive black hole, with estimated mass 
M$_{BH} \sim 10^9$ M$_{\sun}$ (Marchesini et al.~2004),
we can derive a typical distance of $\sim$0.3~pc.
The similarity of the Fe K$\alpha$ line width with that of the broad H$\alpha$ line 
in the optical spectrum (Eracleous \& Halpern 1994) and the distance on sub-pc scales 
possibly suggest an origin of this reflecting material from the BLR.  
Using the distance estimate in the definition of the ionization parameter 
$\xi=L_{ion}/nr^2$ (Tarter, Tucker \& Salpeter 1969), where $n$ is the average absorber
number density and $L_{ion}$ is the source X-ray ionizing luminosity
integrated between 1~Ryd and 1000~Ryd (1~Ryd=13.6~eV), we can derive
the density of the material at that location. 
The absorption corrected ionizing luminosity of \3c\ is 
$L_{ion}$$=$$8\times 10^{44}$~erg~s$^{-1}$. Substituting this value, we can estimate 
the density of the material of $\sim$$5\times 10^7$~cm$^{-3}$.

Moreover, we also detected a
weak emission line at E$\simeq$7.5~keV, most probably ascribable to
 Ni K$\alpha$ fluorescence emission line from the same material
of the Fe K$\alpha$. In particular, the \feka\ is expected to be
more intense than the Ni K$\alpha$ due to the higher cosmic abundance of
iron with respect to nickel, about $\sim$20 times (Molendi, Bianchi \& Matt 
2003; Yaqoob \& Murphy 2011). However, the ratios of the intensities of 
the two emission lines suggest a possible slight overabundance 
of nickel and/or an underabundance of iron of a factor of $\sim$2 with 
respect to the standard solar values (e.g., Anders \& Grevesse 1989; 
Grevesse et al.~1996). Unfortunately, we note that the treatment of Ni
lines is not included in the present version of the \emph{xillver} code (March 2010).
However, the possible slight underabundance of iron or overabundance of Ni has 
negligible effects, within the 1$\sigma$ errors, on the parameters of the mildly
ionized reflection component.

Our most interesting result from the Fe K region modeling is
the detection of a relativistic profile for the broad component of the
\feka\ line, confirming a previous tentative \asca\ result (Reynolds
1997; S99). The \suzaku\ data quality is sufficient to
provide tight constraints on the disk parameters, albeit in a
model-dependent way, in particular on the inner and outer radii of the
emitting region and the disk inclination. We find that the bulk of the
reprocessing in \3c\ occurs in a small disk annulus, $\Delta r \sim 10
r_g$, located at a distance $r_{in} \sim 10 r_g$ from the central
black hole and at an inclination $i$$\sim$25\deg--30\deg.
Substituting the average distance of $r$$\sim$15$r_g$ in the ionization parameter 
equation and the estimated value of log$\xi$$\simeq$3~erg~s$^{-1}$~cm (Table~5), we 
can derive a density for the associated highly ionized reflection component of 
$\sim$$10^{11}$~cm$^{-3}$. This is a realistic estimate for the inner parts of an 
accretion disk in AGN (Garc{\'{\i}}a, Kallman \& Mushotzky 2011).  

A similar finding - that the innermost disk regions do not contribute 
to the Fe~K emission -- was previously reported for other broad-lined 
radio-loud AGN, the BLRGs 3C~120 and 3C~390.3 (Kataoka et al. 2007, 
S09), and the quasar/BLRG 4C+74.26 (Larsson et al.~2008). In these 
sources the inner radius of the Fe~K emission was determined to be 
around 10--50$r_g$ from fits to the \suzaku\ data. Thus, a pattern is 
starting to emerge, whereas the innermost regions of the accretion 
disk in broad-lined radio-loud AGN do not contribute to the Fe~K 
emission. 
  
A number of plausible physical scenarios have been put forth to 
explain the above lack of Fe~K reprocessing in the innermost radii of 
the BLRG disks (see S09 and discussion therein). These include a 
highly ionized ion torus/ADAF occupying the inner disk, obscuration by 
the base of a jet, or lack of illumination of the inner radii of the 
disk by the primary power law, due to a mildly beamed source of 
continuum. An additional possibility is offered in the context of the 
``flux-trap'' scenario (Reynolds et al. 2006; Garofalo 2009). 
  
It is generally believed that jet power is, at least in part, linked
to the spin of the central black hole. Radio-loud AGNs have been
claimed to harbor rapidly spinning black holes (e.g., Nemmen et
al. 2007, and references therein). A scenario for jet power production
is provided by the ``flux trap'' model (Reynolds et al. 2006; Garofalo
2009), which suggests that the plunging region between accretion disks
and black holes is fundamental in producing strong, spin dependent,
horizon-threading magnetic fields, and therefore powerful
jets. Noteworthy, it has been demonstrated that the Blandford \&
Znajek (1977; BZ) mechanisms is maximized for black hole spin $a \sim
-1$ (retrograde) and stronger jets/outflows are expected in this case
(Garofalo 2009). This is a direct consequence of the fact that the
radius of marginal stability for retrograde black holes, $r_{ISCO}\sim
9$r$_g$ for $a=-0.998$, is much larger than in prograde, $r_{ISCO}\sim
1.23$r$_g$ for $a=+0.998$, and therefore the plunging region is more
extendend and the flux-trapping effect is magnified. In principle,
this could be directly tested with X-ray spectroscopy because in the
two cases the relativistic Fe K line profile changes significantly
(Garofalo, Evans, \& Sambruna 2010).
 
As discussed in \S~5.3 and \S~5.4, we tested the retrograde/prograde
scenario by performing spectral fits to the data with a model that
allows the black hole spin to assume {\it negative} values (Dauser et
al. 2010).  While the fit is statistically satisfactory, unfortunately
the spin parameter is unconstrained due to the insufficient S/N ratio
of the data.  However, because of the shape of its Fe~K line,
\3c\ qualifies as the best candidate so far for determining the black
hole spin in a radio-loud AGN through Fe~K line spectroscopy.  Longer
exposures with \suzaku\ are required. 
 
\3c\ shares with two other BLRGs (3C~390.3 and 3C~120) the evidence in 
its X-ray spectrum for emission from an ionized disk, with similar 
ionization parameters $\xi \sim 2300-2700$. This supports the previous 
suggestion by Ballantyne et al.~(2002) that ionized disks play a 
significant role for the X-ray emission of powerful radio-loud AGN. 
Evidence for ionized material in the nuclear regions of radio-loud AGN 
is also provided by the detection of extremely ionized outflows/winds 
with velocities $\ga$10,000~km/s, and properties similar to 
radio-quiet AGN outflows (Tombesi et al.~2010a,b; see also below). 
  
Finally, we compare the X-ray properties of \3c\ and other BLRGs  
observed with \suzaku\ to Seyfert 1s. We started this comparison  
in our previous paper on 3C~390.3 (S09), where we outlined the average  
BLRG properties -- X-ray photon index, Fe~K line widths, and cold  
reflection strengths -- with the then available most current X-ray  
data for Seyferts from \xmm\ EPIC observations (Nandra et  
al. 2007). We noted how the \xmm\ data for radio-quiet AGN suggested a  
much larger spread of reflection fractions and line widths for Seyferts  
than in the \asca\ and \rxte\ era, and how this spread lead to a blur  
in the distinction between radio-loud and radio-quiet sources. In  
particular, in BLRGs X-ray reflection and continuum properties occupy  
one end of the distribution for Seyferts, with significant overlap. We  
concluded that the division between radio-loud and radio-quiet AGN was  
blurred, as far as their X-ray spectral properties were concerned (S09).   
  
A more proper comparison is now possible thanks to the recent
publication of \suzaku\ and \swift\ BAT observations for a relatively
large (6 sources) sample of Seyfert 1s (Patrick et al. 2010). From an
accurate modeling of the 0.6-100~keV spectra using both cold and
ionized reflection models, Patrick et al. (2010) found broad
relativistic profiles for the Fe~K emission lines in most Seyferts,
with average equivalent width for the sample of EW=119$\pm$19 eV and
ionized emission lines at 6.7 and 6.97~keV being relatively
common. Intriguingly, the Fe~K emission seems to arise from tens of
r$_g$ from the central black hole (see also Nandra et al. 2007); fits
with models allowing the black hole spin to vary, although in the 0--1
range only, yield typically intermediate values, $a\sim 0.7$. Both
cold and ionized reflection is required to describe the data above
10~keV; the cold reflection fraction, however, varies from source to
source from less than 0.4 up to 2.5 (see Table~4 in Patrick et
al. 2010). The intrinsic X-ray photon index is $\Gamma_X \sim
1.8-2.2$, consistent with Comptonization models. However, 
the ionization parameter of the ionized disk reflection seems to be
lower in the Seyferts, consistent with neutral or lowly ionized Fe,
while interestingly so far the BLRGs seem to suggest high ionization
disk lines (He and H-like iron), although based on small number
statistics so far.
  
Thus, comparison of the {\it broad-band 0.6--200~keV} spectra of BLRGs 
and Seyfert 1s obtained with \suzaku\ and the BAT confirm and 
reinforce the conclusion (S09) that there is a continuum of X-ray 
properties between the two AGN subclasses, with radio-loud sources 
clustered at one end of the distribution. An important consideration, 
however, is that in both subclasses -- aside for a few egregious 
examples -- the innermost $\sim 10$r$_g$ of the accretion disk do not 
contribute significantly to Fe~K emission. The reason for this may be 
different in the two subclasses, and would remain of central importance 
for the radio-loud/radio-quiet AGN division. 
  
Another major progress since S09 is the discovery that radio-loud AGNs 
have ionized, outflowing material from the central nuclei, similar to 
Seyferts (e.g., Tombesi et al.~2010a,b), contrary to what stated in 
S09 based on the then available evidence. Indeed, our 
\suzaku\ observations of 5 BLRGs provided evidence for ultra-fast disk 
outflows in 3/5 sources (Tombesi et al.~2010b) with velocities 
$v_{out} \sim 0.1$c, kinetic power $\sim 10^{43-44}$ \lum, and mass 
outflow rates comparable to the accretion rates. As mentioned above 
and in \S~2, \3c\ itself presents evidence for a highly ionized, 
kpc-scale outflow through a series of soft X-ray absorption features 
detected with the \chandra\ and \xmm\ gratings (Reeves et al. 2009; 
Torresi et al. 2010), although with a much lower velocity of $\sim$1,000~km/s.
 The presence of ionized material along the line 
of sight is puzzling and its relation with the relativistic jet must 
be studied in much detail. As already discussed in Tombesi et 
al.~(2010b), we did not detect any significant highly ionized Fe~K 
absorption line in the present observation of \3c.  
  
The overlapping X-ray continuum and reflection properties of BLRGs and
Seyferts point to a common accretion flow structure, with similar
radiative processes in place that account broadly for the X-ray
emission. The reason why powerful, relativistic jets are present in
BLRGs but {\it not} in Seyferts with similar engines could very well
be identified in a process unrelated to the accretion dynamics --
i.e., the black hole spin. Several theoretical studies along this line
are underway, and the wealth of high-quality data that
\suzaku\ continues to provide is essential to test the model
predictions in the X-ray band, paving the way for the upcoming
high-precision spectroscopy with the {\it Astro-H} observatory.

\acknowledgements   
 
This research has made use of data obtained from the High Energy 
Astrophysics Science Archive Research Center (HEASARC), provided by 
NASA's Goddard Space Flight Center, and of the NASA/IPAC Extragalactic 
Database (NED) which is operated by the Jet Propulsion Laboratory, 
California Institute of Technology, under contract with the National 
Aeronautics and Space Administration. RMS and FT acknowledge support 
from NASA through the \suzaku\ and ADAP/LTSA programs. FT thank T. Dauser 
for help in using the \emph{relline} model. FT thank J. Garc{\'{\i}}a for 
help in using the \emph{xillver} model. LB acknowledges financial support from 
the Spanish Ministry of Science and Innovation through a ``Juan de la Cierva'' 
fellowship, research grant AYA2009-08059. The authors thank the referee
for suggestions that led to important improvements in the paper. 
  
\appendix  
\section{Background and Calibration tests for the XIS}

We performed detailed background and calibration tests for the three  
XIS detectors to exclude possible issues affecting the Fe~K region,  
and thus the results of our modeling described above. Background  
contamination could alter the profile of the Fe line;  
however, for the \3c\ observation analyzed here, we estimate that the  
background contribution to 0.6--10~keV is negligible, amounting to  
$\la$4--8\% of the source counts.   
  
It is well known that the XIS CCDs contain a $^{55}$Fe calibration  
source, located on two corners of the XIS chips, which produces a  
characteristic X-ray line from Mn~K$\alpha$ at E$=$5.895~keV. Its  
intrinsic width is $\sigma$$\la$50~eV and it is used to calibrate the  
energy and width of the iron line (Koyama et al.~2007).  We can  
clearly exclude any possible significant contamination of the broad  
line detected at the rest-frame energy of E$\simeq$5.8~keV from this  
calibration source on board the XIS.  In fact, we checked the images  
of the three separate XIS cameras and find that the source is located  
in the center of the CCD, far from the calibration sources, which are in two  
corners.  Moreover, the observed energy of the broad line is at  
E$=$5.45~keV, which is not consistent with that expected from the  
calibration source, and the calibration line is intrinsically  
narrower compared with the $\simeq$400~eV of the \3c\ feature.  
  
As already discussed in \S~3.1 the combined XIS-FI and XIS-BI fit is supported by the 
consistency of the power-law slope and 2--10~keV flux within $\sim$2\%.
Moreover, as already reported in \S~4, we checked that the presence of two low amplitude 
($\sim$5\%) spikes in the XIS~3 light curve, see Figure~1, have negligible effects on the 
overall spectral analysis.  
This is expected from the fact that we focused on the average spectrum and the weak short 
time-scale spikes involved less than 20\% of the exposure.  
 These are most probably ascribable to weak instrumental fluctuations.
We directly analyzed the three XIS cameras separately and ensured that there is 
an agreement at $\sim$1\% for both the power-law continuum slope and 2--10~keV flux between 
the XIS~0 and XIS~3 and at $\sim$2\% between the front illuminated XIS~0-3 and back 
illuminated XIS~1.

Finally, the spectra of the individual XIS camera were fitted  
separately for a consistency check of the emission line  
parameters. Unfortunately, the quality of the individual camera spectra  
does not allow us to leave all line parameters free to vary during  
the fit. Using the best-fit model determined from the analysis of the  
joint spectra (see Table~3), and freezing the lines energies and widths, 
we find that the EWs of the emission lines are always consistent at 90\% among  
the separate XIS instruments and the combined XIS-FI and XIS-BI fit.

\clearpage


\begin{deluxetable}{cccccc}
\tabletypesize{\footnotesize}
\tablecaption{3C~382 observations log.}
\tablewidth{0pt}
\tablehead{\colhead{Satellite} & \colhead{OBSID} & \colhead{Date} & \colhead{Instrument} & \colhead{Net expo} & \colhead{Rate}\\ & & & & \colhead{(ks)} & \colhead{(cts/s)}}
\startdata
Suzaku & 702125010 & 2007 April 27 & XIS-FI & 116 & 2.399/1.219\\[2.5pt]
       &           &               & XIS-BI & 116 & 1.613/1.177\\[2.5pt]
       &           &               & PIN    & 113 & 0.114\\[2.5pt]
XMM-Newton & 0506120101 & 2008 April 28 & EPIC pn & 21 & 15.650/4.220\\[2.5pt]
Swift & J1835.0$+$3240 & & BAT & 4276 & 0.001\\[2.5pt]
\enddata
\tablecomments{The count rates refer to each instrument separately and are in the 0.5--2/2--10~keV band for XIS-FI, XIS-BI and EPIC pn, 14--70~keV band for the PIN and 15--100~keV band for the BAT.}

\end{deluxetable}   
\begin{deluxetable}{c|cc|cc}
\tabletypesize{\footnotesize}
\tablecaption{Flux and luminosities from Suzaku and XMM.}
\tablewidth{0pt}
\tablehead{ Energy & \multicolumn{2}{c}{Suzaku} & \multicolumn{2}{c}{XMM}\\ 
 (keV) & Flux & Lum & Flux & Lum}
\startdata
0.5--2 & $2.42 (3.10) \pm 0.02$ & $1.92 (2.51) \pm 0.02$ & $2.59 (3.44) \pm 0.02$ & $2.05 (2.79) \pm 0.01$\\[2.5pt]
2--10 & $4.09 (4.12) \pm 0.01$ & $3.25 (3.28) \pm 0.01$ & $3.83 (3.88) \pm 0.02$ & $3.05 (3.09) \pm 0.01$\\[2.5pt] 
15--200 & $8.10 (8.10) \pm 0.40$ & $6.50 (6.53) \pm 0.30$ & & \\[2.5pt]
\enddata
\tablecomments{The observed (un-absorbed) flux is in units of $10^{-11}$~erg~s$^{-1}$~cm$^{-2}$. The observed (un-absorbed) luminosity is in units of $10^{44}$~erg/s.}

\end{deluxetable}   
\begin{deluxetable}{ccccc}
\tabletypesize{\scriptsize}
\tablecaption{Best-fit of the E$=$0.6-200~keV broad-band spectrum.}
\tablewidth{0pt}
\tablehead{\colhead{Pexrav} & & & &}
\startdata
$\Gamma$ & E$_c$ & $R$ & $i$ & $\chi^2/\nu$\\
 & (keV) & & (deg) &\\ 
\hline
$1.859\pm0.004$ & $304^{+107}_{-95}$ & $0.36\pm0.06$ & 30 & 4416/4167\\
\hline\hline\\[-4pt]
Warm absorber & & & &\\
\hline\\[-4pt]
N$_H$ & log$\xi$ & $v_{out}$ & &\\
($10^{20}$~cm$^{-2}$) & (erg~s$^{-1}$~cm$^{-2}$) & (km/s) & &\\ 
\hline
 $5.8\pm1.5$ & $2.44^{+0.07}_{-0.04}$ & 1000 & &\\
\hline\hline\\[-4pt]
Black body & & & &\\
\hline\\[-4pt]
$k$T & Norm & & &\\
 (eV) & ($10^{-4}$)\tablenotemark{a} & & &\\
\hline
$100\pm4$ & $1.4\pm0.2$ & & &\\
\hline\hline\\[-4pt]
Gaussians & & & &\\
\hline\\[-4pt]
E & $\sigma$ & $I$ & EW & ID\\
(keV) & (eV) & ($10^{-5}$~ph~s$^{-1}$~cm$^{-2}$) & (eV) &\\ 
\hline
$6.41(6.06)\pm0.01$ & $118^{+20}_{-13}$ & $3.3\pm0.2$ & $68\pm9$ & Fe K$\alpha$\\
$\equiv7.06(6.67)$ & & & $10$ & Fe K$\beta$\\
$7.51(7.10)\pm0.03$ & $\equiv 10$ & $0.7\pm0.1$ & $18\pm7$ & Ni K$\alpha$\\
$5.70(5.39)\pm0.09$ & $385^{+77}_{-57}$ & $3.0\pm0.4$ & $55\pm12$ &\\
$6.91(6.53)\pm0.02$ & $\equiv 10$ & $0.7\pm0.1$ & $17\pm6$ &\\
\enddata
\tablecomments{Energy of the line, rest-frame (observed frame). Errors are at the 1$\sigma$ level.}
\tablenotetext{a}{Normalization of the blackbody model in units of $L_{39}/D^2_{10}$, where $L_{39}$ is the source luminosity in $10^{39}$~erg~s$^{-1}$ and $D_{10}$ the distance of the source in units of 10~kpc.}
\end{deluxetable}
   
\begin{deluxetable}{ccccc}
\tabletypesize{\scriptsize}
\tablecaption{Best-fit of the E$=$0.6-200~keV broad-band spectrum, including the ionized reflection component.}
\tablewidth{0pt}
\tablehead{\colhead{Pexrav} & & & &}
\startdata
$\Gamma$ & E$_c$ & $R$ & $i$ & $\chi^2/\nu$\\
 & (keV) & & (deg) &\\ 
\hline
$1.741\pm0.003$ & $187^{+44}_{-31}$ & $0.15\pm0.05$ & 30 & 4430/4168\\
\hline\hline\\[-4pt]
Warm absorber & & & &\\
\hline\\[-4pt]
N$_H$ & log$\xi$ & $v_{out}$ & &\\
($10^{20}$~cm$^{-2}$) & (erg~s$^{-1}$~cm$^{-2}$) & (km/s) & &\\ 
\hline
 $8.5\pm1.5$ & $2.65^{+0.04}_{-0.06}$ & 1000 & &\\
\hline\hline\\[-4pt]
Black body & & & &\\
\hline\\[-4pt]
$k$T & Norm & & &\\
 (eV) & ($10^{-4}$)\tablenotemark{a} & & &\\
\hline
$104\pm3$ & $1.4\pm0.5$ & & &\\
\hline\hline\\[-4pt]
Gaussians & & & &\\
\hline\\[-4pt]
E & $\sigma$ & $I$ & EW & ID\\
(keV) & (eV) & ($10^{-5}$~ph~s$^{-1}$~cm$^{-2}$) & (eV) &\\ 
\hline
$6.41(6.06)\pm0.01$ & $86^{+12}_{-15}$ & $2.1\pm0.2$ & $42\pm6$ & Fe K$\alpha$\\
$\equiv7.06(6.67)$ & & & $6$ & Fe K$\beta$\\
$7.51(7.1)\pm0.03$ & $\equiv 10$ & $0.5\pm0.1$ & $13\pm7$ & Ni K$\alpha$\\
\hline\hline\\[-4pt]
Relconv & & & &\\
\hline\\[-4pt]
$\beta$ & $a$ & $r_{out}$ & $i$ &\\
&  & ($r_g$) & (deg) &\\ 
\hline
 $-2.5$ & $-0.998$ & $26^{+6}_{-4}$ & $29\pm2$ &\\
\hline\hline\\[-4pt]
Xillver & & & &\\
\hline\\[-4pt]
log$\xi$ & Norm & & &\\
 (erg~s$^{-1}$~cm) & ($10^{-8}$) & & &\\
\hline
$3.04^{+0.03}_{-0.04}$ & $4.7^{+0.5}_{-0.3}$ & & &\\
\enddata
\tablecomments{Energy of the line, rest-frame (observed frame). Errors are at the 1$\sigma$ level.}
\tablenotetext{a}{Normalization of the blackbody model in units of $L_{39}/D^2_{10}$, where $L_{39}$ is the source luminosity in $10^{39}$~erg~s$^{-1}$ and $D_{10}$ the distance of the source in units of 10~kpc.}
\end{deluxetable}
   
\begin{deluxetable}{cccc}
\tabletypesize{\scriptsize}
\tablecaption{Best-fit of the E$=$0.6-200~keV broad-band spectrum, including two ionized reflection components.}
\tablewidth{0pt}
\tablehead{\colhead{Cutoff power-law} & & &}
\startdata
$\Gamma$ & E$_c$ & & $\chi^2/\nu$\\
 & (keV) & &\\ 
\hline
$1.737\pm0.003$ & $175^{+25}_{-20}$ & & 4467/4174\\
\hline\hline\\[-4pt]
Warm absorber & & &\\
\hline\\[-4pt]
N$_H$ & log$\xi$ & $v_{out}$ &\\
($10^{20}$~cm$^{-2}$) & (erg~s$^{-1}$~cm$^{-2}$) & (km/s) &\\ 
\hline
 $8.3\pm1.5$ & $2.51^{+0.05}_{-0.06}$ & 1000 &\\
\hline\hline\\[-4pt]
\multicolumn{4}{l}{Xillver (mildly ionized)}\\
\hline\\[-4pt]
log$\xi$ & Norm & $R_F$ &\\
 (erg~s$^{-1}$~cm) & ($10^{-6}$) & &\\
\hline
$1.54\pm0.03$ & $3.0\pm0.3$ & 0.10 &\\
\hline\hline\\[-4pt]
Relconv & & &\\
\hline\\[-4pt]
$\beta$ & $a$ & $r_{out}$ & $i$\\
&  & ($r_g$) & (deg)\\ 
\hline
 $-2.5$ & $-0.998$ & $25^{+6}_{-5}$ & $30\pm3$\\
\hline\hline\\[-4pt]
\multicolumn{4}{l}{Xillver (highly ionized)}\\
\hline\\[-4pt]
log$\xi$ & Norm & $R_F$ &\\
 (erg~s$^{-1}$~cm) & ($10^{-8}$) & &\\
\hline
$2.93\pm0.04$ & $4.9^{+0.4}_{-0.3}$ & 0.11 &\\
\enddata
\tablecomments{Energy of the line, rest-frame (observed frame). Errors are at the 1$\sigma$ level.}
\tablenotetext{a}{Normalization of the blackbody model in units of $L_{39}/D^2_{10}$, where $L_{39}$ is the source luminosity in $10^{39}$~erg~s$^{-1}$ and $D_{10}$ the distance of the source in units of 10~kpc.}
\end{deluxetable}
   

\clearpage  
\newpage


   \begin{figure}[ht]  
   \centering  
    \includegraphics[width=8cm,height=9.5cm,angle=0]{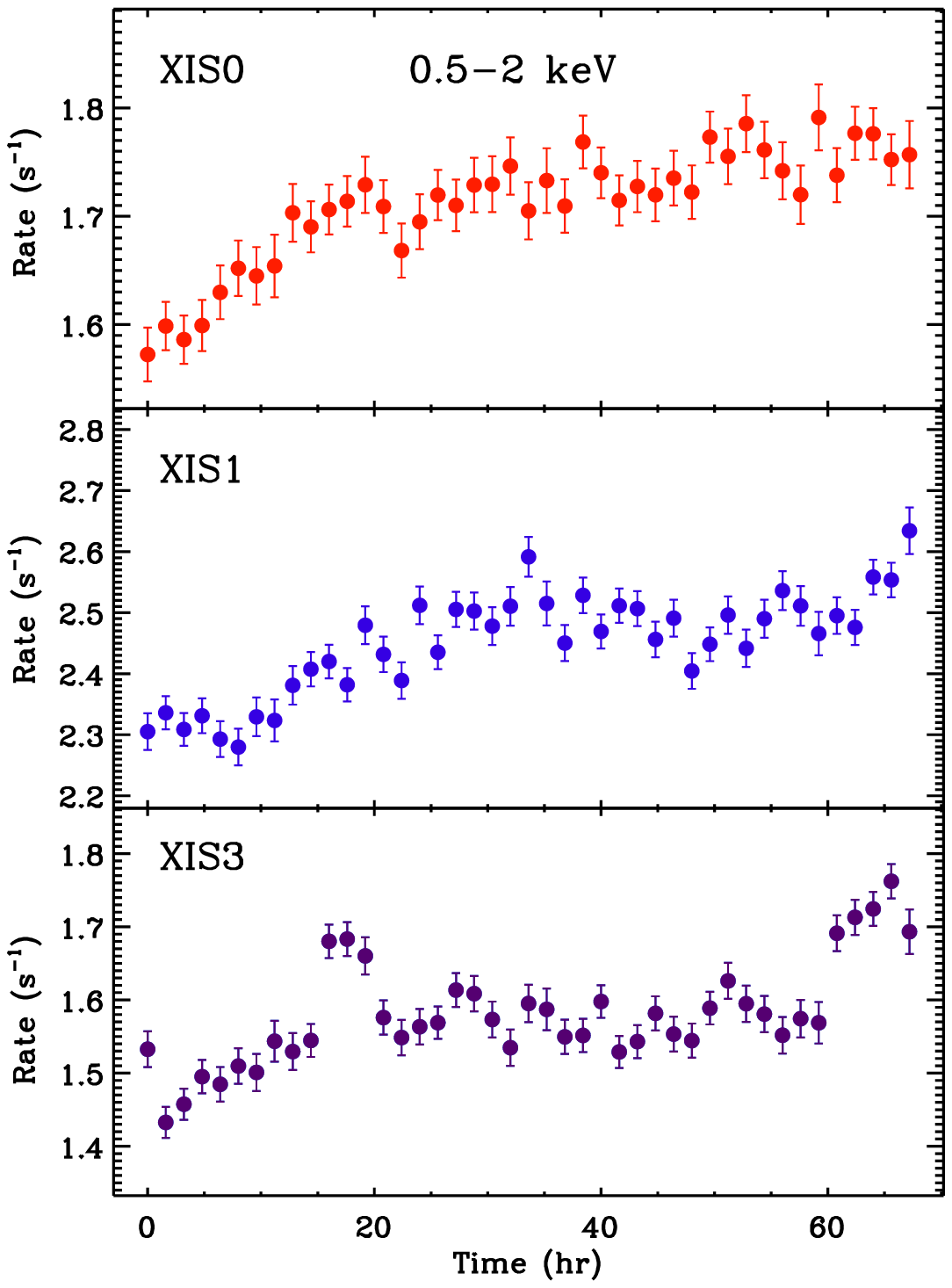}
    \hspace{-2cm}
    \includegraphics[width=8cm,height=9.5cm,angle=0]{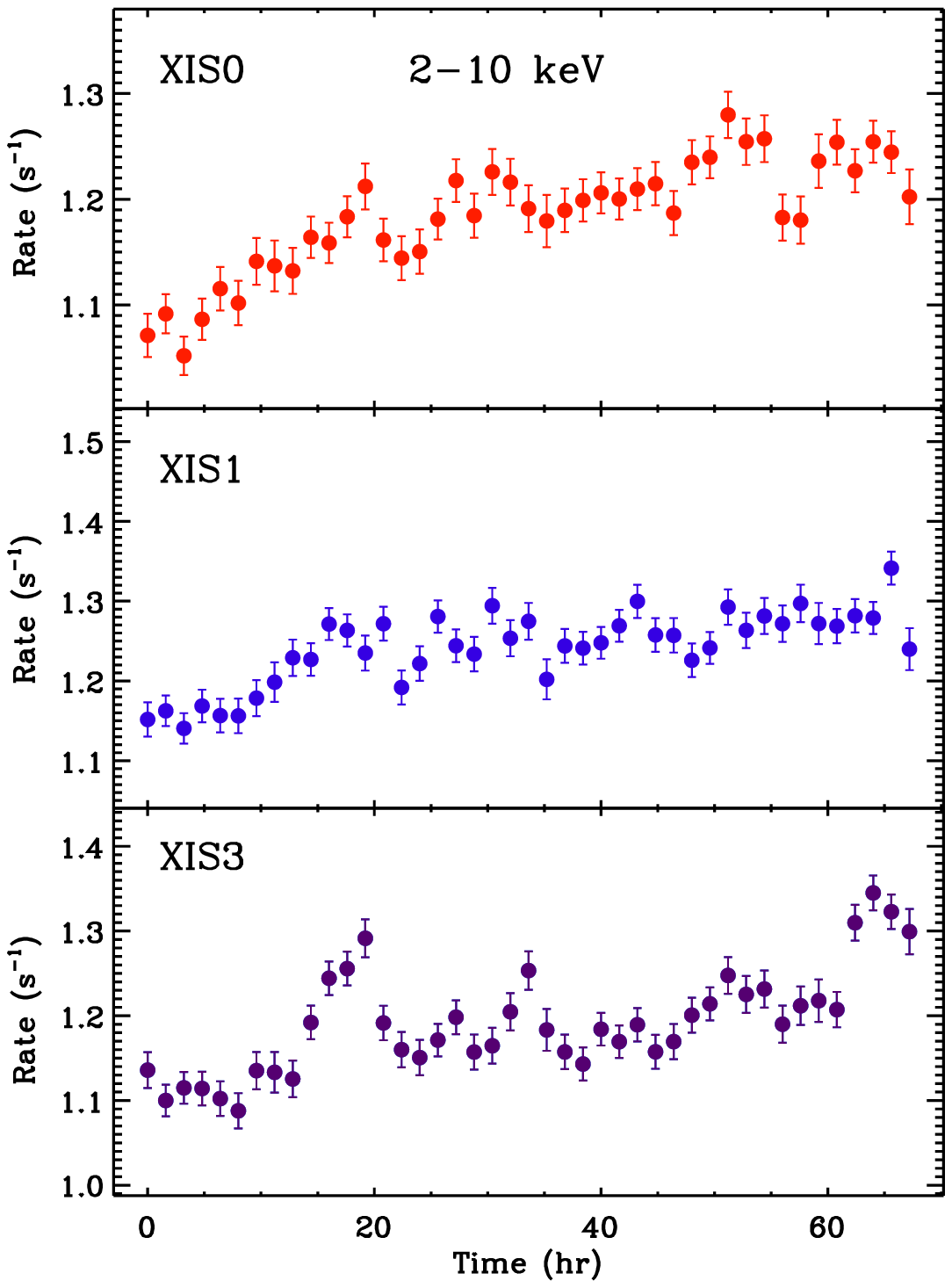}
   \caption{Light curves of \3c\ at soft (a, left) and hard (b, right)  
     X-rays from the \suzaku\ XIS0-3 observations. The data from the  
     three detectors are plotted in separate panels for clarity. Bins  
     are 5760~sec, or 1 orbit.}  
    \end{figure}


   \begin{figure}[ht]  
   \centering  
    \includegraphics[width=8cm,height=12cm,angle=0]{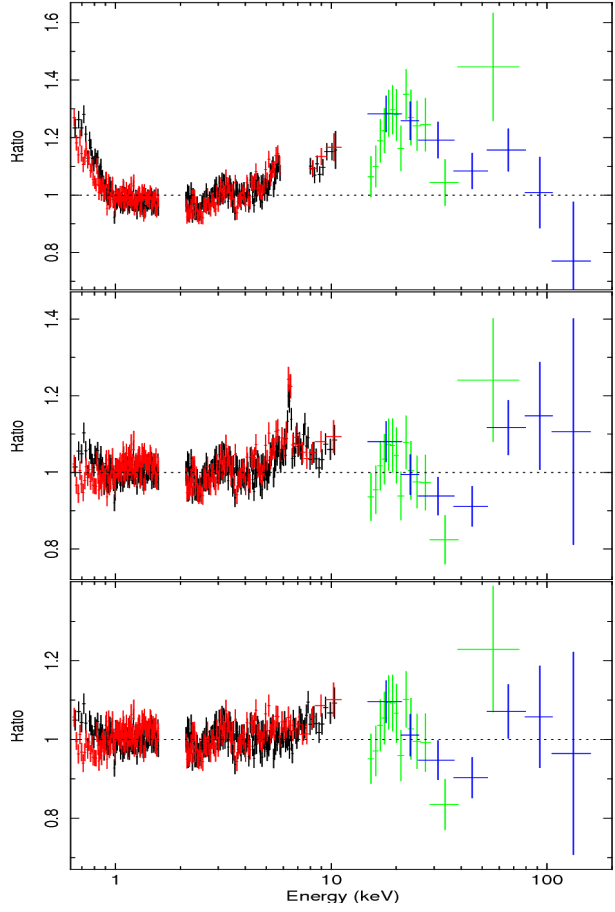}  
   \caption{Combined XIS-FI (black), XIS-BI (red), PIN  
       (green) and BAT (blue) broad-band spectra in the  
       E$=$0.6--200~keV energy band. Data have been rebinned to  
       S/N$=$40 for XIS and S/N$=$12 for PIN only for plotting. 
       \emph{Upper panel:} ratios with respect to a Galactic 
       absorbed power-law continuum. The Fe K band between 5.5--7.5~keV is 
       excluded. \emph{Middle panel:} ratios after the inclusion of a 
       blackbody soft excess and a cold reflection component. \emph{Lower 
       panel:} ratios after the inclusion of a warm absorber and five 
       Gaussian emission lines in the Fe K band.}  
    \end{figure}


   \begin{figure}[ht]  
   \centering  
    \includegraphics[width=8cm,height=12cm,angle=0]{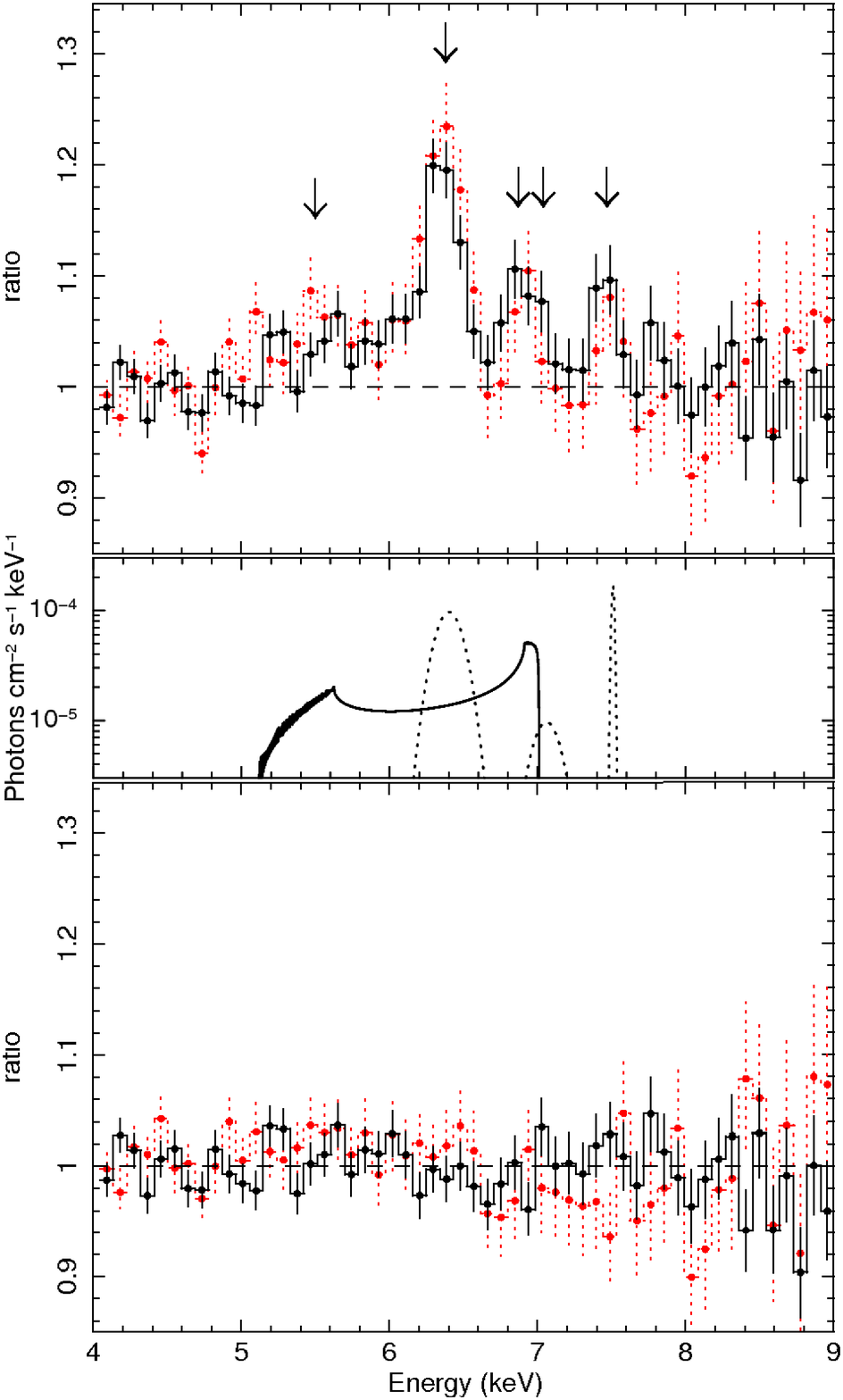}  
   \caption{Zoomed spectrum of 3C~382 in the Fe K band, E$=$4--9~keV. 
        The XIS-FI (black solid line) and XIS-BI (red dotted line) have been both rebinned 
        to 100~eV energy bins only for plotting.
       \emph{Upper panel:} data/model ratios with respect to the 
       best-fit continuum model without the inclusion of emission lines in 
       the Fe K band. The arrows indicate the detected Gaussian emission 
       lines.  \emph{Middle panel:} best-fit model composed by three
       narrow Gausssian emission lines and a \emph{diskline} profile. 
       \emph{Lower panel:} data/model ratios with respect to the model 
       including three Gaussian emission lines and a \emph{diskline}.} 
    \end{figure}


   \begin{figure}[ht]  
   \centering  
    \includegraphics[width=8cm,height=9cm,angle=0]{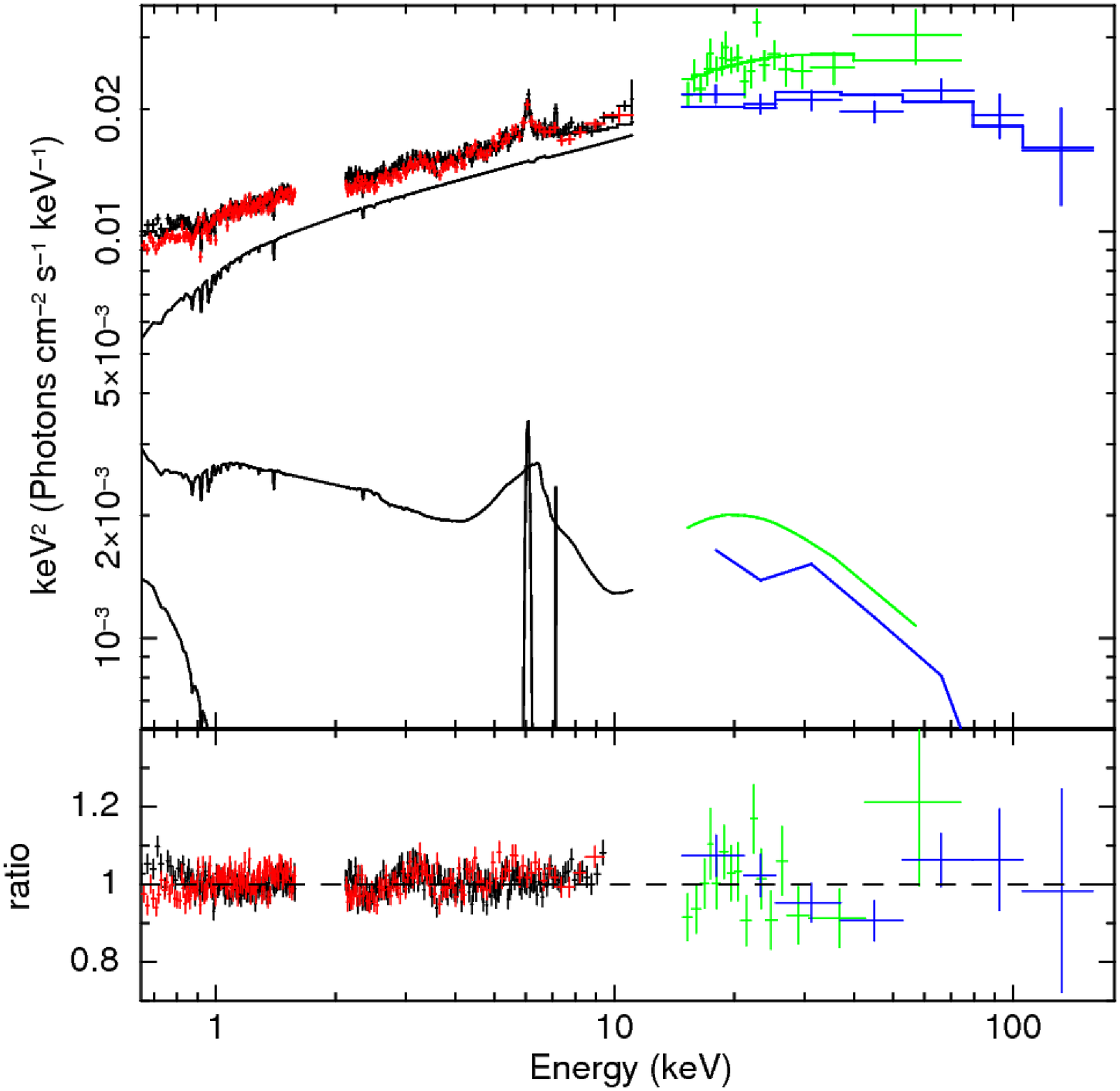} 
    \includegraphics[width=8cm,height=9cm,angle=0]{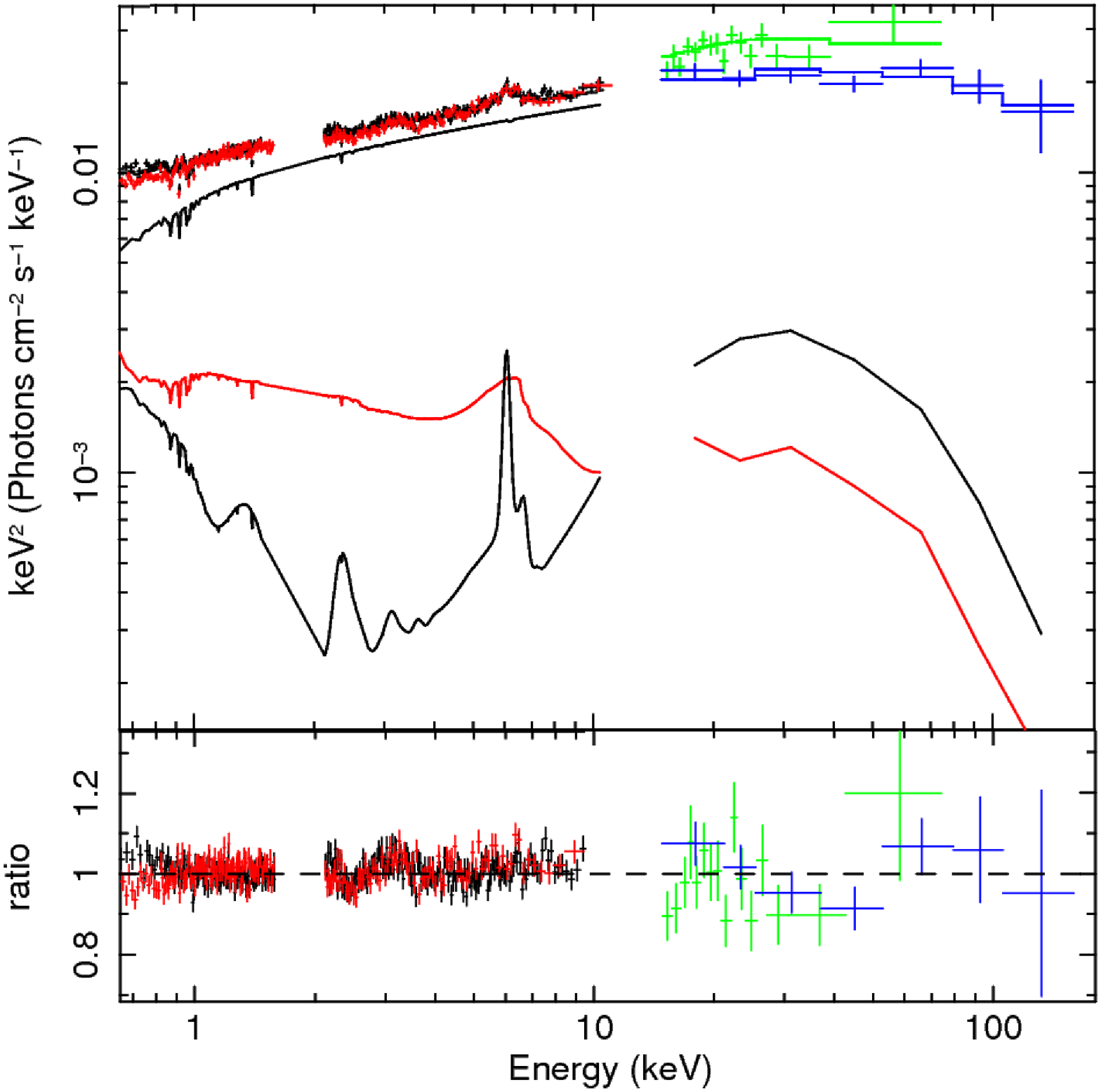}   
   \caption{Combined XIS-FI (black), XIS-BI (red), PIN  
       (green) and BAT (blue) broad-band spectra in the  
       E$=$0.6--200~keV energy band. Data have been rebinned to  
       S/N$=$40 for XIS and S/N$=$12 for PIN only for plotting.
       \emph{Left:} broad-band model including the black body soft excess, 
       Gaussian emission lines and cold \emph{pexrav} reflection component, see \S~5.4 
       and Table~4. \emph{Right:} broad-band model including only two ionized 
       reflection components, see \S~5.5 and Table~5. \emph{Upper panels:} data 
       with best-fit model.  \emph{Lower panels:} ratio with  
       respect to the best-fit model.} 
    \end{figure}  
  

   \begin{figure}[ht]  
   \centering  
    \includegraphics[width=13cm,height=10cm,angle=0]{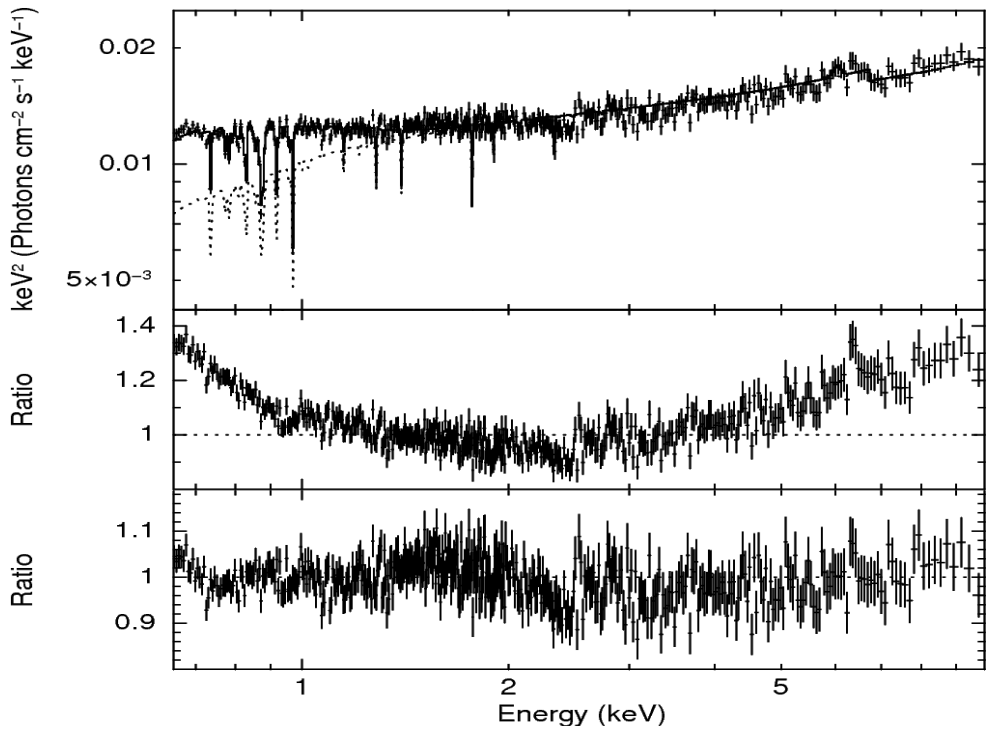}  
   \caption{\xmm\ EPIC pn spectrum of 3C~382 in  
       the E$=$0.6--10~keV band. Data have been rebinned to a S/N of 20 only 
       for plotting. \emph{Upper panel:} data and best-fit  
       model. \emph{Middle panel:} ratio against a Galactic absorbed  
       power-law continuum. \emph{Lower panel:} ratio against the best-fit 
       model.}  
    \end{figure}

\end{document}